\documentclass[11pt,a4paper]{article}
\usepackage{fancyhdr} 
\usepackage{amsmath} \usepackage{amsfonts} \usepackage{amssymb}
\usepackage{wasysym}
\usepackage{xcolor}
\usepackage{graphicx}
\usepackage{multirow}
\usepackage[hang,footnotesize]{caption}
\usepackage{epstopdf}

\renewcommand{\d}[1]{{\mathrm d}#1}
\newcommand{\fn}[2]{\mathinner{#1\mathopen{\left(#2\right)}}}
\newcommand{\erf}{\mathop{\mathrm{erf}}}
\newcommand{\erfc}{\mathop{\mathrm{erfc}}}

\newcommand{\remove}[1]{{}}
\newcommand{\Mpc}{\mathinner{\mathrm{Mpc}}}
\newcommand{\Gyr}{\mathinner{\mathrm{Gyr}}}
\newcommand{\typicality}[1]{\fn{\mathcal{T}_+}{#1}}
\newcommand{\eq}[1]{Eq.~(\ref{#1})}

\newcommand{\twoeq}[2]{Eqs.~({\ref{#1}}) and ({\ref{#2}})}
\newcommand{\fig}[1]{Figure~\ref{#1}}
\newcommand{\figs}[2]{Figures~\ref{#1}--\ref{#2}}

\newcommand{\sect}[1]{Section~\ref{#1}}

\newcommand{\tab}[1]{Table~\ref{#1}}

\newcommand{\twotab}[2]{Tables~\ref{#1} and \ref{#2}}

\begin{document}
\author{
Sungwook~E.~Hong${}^{1, 2}$\thanks{eostm@muon.kaist.ac.kr},
Ewan~D.~Stewart${}^{1, 3}$ and
Heeseung~Zoe${}^{1, 4}$\thanks{heezoe@gmail.com} 
\medskip \\
\normalsize
\emph{ ${}^1$Department of Physics, KAIST} \\
\normalsize
\emph{Daejeon 305-701, Republic of Korea}
\\
\normalsize
\emph{ ${}^2$Department of Astronomy and Space Science, Chungnam National University} \\
\normalsize
\emph{Daejeon 305-764, Korea}
\\
\normalsize
\emph{ ${}^3$Yukawa Institute for Theoretical Physics, Kyoto University} \\
\normalsize
\emph{Kyoto 606-8502, Japan}
\\
\normalsize
\emph{${}^4$Department of Physics, Middle East Technical University} \\
\normalsize
\emph{Ankara 06531, Turkey}
}
\title{Anthropic Likelihood for the Cosmological Constant and the Primordial Density Perturbation Amplitude}
\date{October 14, 2011}
\maketitle
\begin{abstract}
Weinberg et al.\ calculated the anthropic likelihood of the cosmological constant $\Lambda$
using a model assuming that the number of observers is proportional to the total mass of gravitationally collapsed objects,
with mass greater than a certain threshold, at $t \to \infty$.
We argue that Weinberg's model is biased toward small $\Lambda$,
and to try to avoid this bias we modify his model
in a way that the number of observers is proportional to the number of collapsed objects,
with mass and time equal to certain preferred mass and time scales.
The Press-Schechter formalism, which we use to count the collapsed objects, identifies our collapsed object at the present time as the Local Group, making it inconsistent to choose the preferred mass scale as that of the Milky Way at the present time.
Instead, we choose an earlier time before the formation of the Local Group
and this makes it consistent to choose the mass scale as that of the Milky Way.
Compared to Weinberg's model ($\typicality{\Lambda_0} \sim 23\%$), 
this model gives a lower anthropic likelihood of $\Lambda_0$ ($\typicality{\Lambda_0} \sim 5\%$).
On the other hand, the anthropic likelihood of the primordial density perturbation amplitude $Q_0$ from this model is high ($\typicality{Q_0} \sim 63\%$),
while the likelihood from Weinberg's model is low ($\typicality{Q_0} \ll 0.1\%$).

Furthermore, observers will be affected by the history of the collapsed object, 
and we introduce a method to calculate the anthropic likelihoods of $\Lambda$ and $Q$ from the mass history using the extended Press-Schechter formalism. 
The anthropic likelihoods for $\Lambda$ and $Q$ from this method are similar to those from our single mass constraint model,
but, unlike models using the single mass constraint which always have degeneracies between $\Lambda$ and $Q$,
the results from models using the mass history are robust even if we allow both $\Lambda$ and $Q$ to vary.

In the case of Weinberg's flat prior distribution of $\Lambda$ (pocket based multiverse measure),
our mass history model gives $\typicality{\Lambda_0} \sim 10\%$,
while the scale factor cutoff measure and the causal patch measure give $\typicality{\Lambda_0} \gtrsim 30\%$.
\end{abstract}

\thispagestyle{fancy}
\lhead{}
\rhead{YITP-11-89}
\newpage
\section{Introduction}
\label{sec:intro}
The observed value of the cosmological constant, $\Lambda_0$, is extremely small, smaller than naive theoretical expectations by a factor of $10^{-60}$ to $10^{-120}$.
Also, this tiny vacuum energy density is \emph{now} of the same order as the matter density.
These two mysteries are the cosmological constant problems, the former is the cosmological constant hierarchy problem, and the later is the cosmological constant coincidence problem.

The most promising solution \cite{Weinberg:1987dv,Weinberg:1988cp,Efstathiou:1995ne,Martel:1997vi,Garriga:1999hu,Garriga:2000cv,Pogosian:2006fx}
to these problems is using anthropic selection \cite{carter,davies,carter_mccrea,barrow_tipler,greenstein,Stewart:2000vu}, 
which notes that we should take into account our own existence when we consider quantities that we observe.
In particular, the probability of observing a given value of the cosmological constant is
\begin{equation}\label{eq:prob}
\fn{P}{\Lambda|\smiley} = \frac{\fn{P}{\Lambda} \fn{P}{\smiley|\Lambda}}{\fn{P}{\smiley}}
\end{equation}
where $\fn{P}{\Lambda}$ is the probability distribution of the cosmological constant in the whole universe and $\fn{P}{\smiley|\Lambda}$ is the probability of finding an observer in a region with cosmological constant $\Lambda$.
Thus, even if $\fn{P}{\Lambda}$ is small, $\fn{P}{\Lambda|\smiley}$ may be large depending on the anthropic likelihood
\begin{equation}
\fn{L}{\Lambda|\smiley} = \frac{\fn{P}{\smiley|\Lambda}}{\fn{P}{\smiley}} .
\end{equation}

For anthropic selection to be able to select the observed value, the observed value must exist.
For it to exist naturally, two things are necessary: a sufficient number of different low energy laws of physics (i.e.\ vacua) to allow the natural existence of the observed value in the laws of physics, and the realization of those low energy laws of physics in different regions of the universe.
String theory calculations have supported the anthropic prediction of at least $10^{10^2}$ different vacua, while both the many-worlds interpretation of quantum mechanics and eternal inflation generate a multiverse realizing the vacua.

To determine $\fn{P}{\Lambda}$ it is necessary to understand the fundamental theory well, much better than our current understanding, and even if this is done, a proper measure for the eternally inflating multiverse is lacking.
Though it seems reasonable to take $\fn{P}{\Lambda}$ to be constant over the anthropically interesting range of the cosmological constant \cite{Weinberg:1987dv,Efstathiou:1995ne},
a choice of cutoff for the eternally inflating multiverse may then modify this flat prior, though it is unknown which, if any, is the correct choice.
In this paper we consider three types of multiverse measure: 
the pocket based measure which Weinberg and Vilenkin assumed \cite{Garriga:1998px,Vanchurin:1999iv,Garriga:2005av,Easther:2005wi}, the scale factor cutoff measure \cite{DeSimone:2008bq,Bousso:2008hz} and the causal patch measure \cite{Bousso:2006ev,Bousso:2006ge}.

To estimate the anthropic likelihood it is necessary to have an anthropic model which relates the number of observers to some calculable quantity.
Weinberg et al.\ \cite{Martel:1997vi} and Vilenkin et al.\ \cite{Pogosian:2006fx} modeled the number of observers as proportional to the total mass in gravitationally collapsed objects with mass greater than a certain threshold, usually taken to be the mass of the Milky Way, at late times.
Using this model they postdicted the observed value of the cosmological constant with an error of 1 to 2$\sigma$.

The total mass in gravitationally collapsed objects
depends not only on the cosmological constant but also on the primordial density perturbation amplitude, $Q$,
and there have been some studies to understand our observed value of the primordial density perturbation amplitude, $Q_0$,
using anthropic selection \cite{Tegmark:1997in,Banks:2003es, Graesser:2004ng,Garriga:2005ee}.
Tegmark \& Rees \cite{Tegmark:1997in}
showed that both too high and too low a primordial density perturbation amplitude may be harmful for observers,
and constrained anthropically allowed values of the primordial density perturbation amplitude to within an order of magnitude of the observed value.

The plan of our paper is as follows.
In \sect{sec:prior}, we review multiverse measures and how they affect the prior.
In \sect{sec:fin}, we review Weinberg's anthropic model \cite{Martel:1997vi,Garriga:1999hu,Garriga:2000cv,Pogosian:2006fx}, discuss its deficiencies, 
and introduce some improved models.
In \sect{sec:his}, we introduce anthropic models using the mass history of the collapsed object.
We summarize our result in \sect{sec:sum} and discuss future work in \sect{sec:dis}.

\section{Prior distribution and choice of multiverse measure}
\label{sec:prior}

As in \eq{eq:prob}, the probability of observing an observable $O$ is
\begin{equation}\label{eq:prior_vague}
\fn{P}{O|\smiley} = \fn{P}{O} \fn{L}{O|\smiley}.
\end{equation}
In this paper we focus on the anthropic likelihood $\fn{L}{O|\smiley}$,
but the prior distribution also affects the probability $\fn{P}{O|\smiley}$,
so we will briefly review possible prior distributions of the cosmological constant and the primordial density perturbation amplitude.

In the eternally inflating multiverse,
the number of observers in each universe is infinite,
so it is ill-defined how to compare the numbers of observers in different types of universe.
However, the comoving anthropic likelihood,
which counts the number of observers in a comoving volume, $\fn{L_\mathrm{c}}{O|\smiley}$, 
is well-defined since all the ambiguity is left in the corresponding prior distribution $\fn{P_\mathrm{c}}{O}$.
$\fn{P_\mathrm{c}}{O}$ can be divided into two parts:
the primordial prior distribution $\fn{P_\varnothing}{O} \equiv \fn{P_\mathrm{c}}{O, t=0}$ which comes from both the fundamental theory and the multiverse ambiguity at the primordial stage,
and an additional factor $\fn{W_\mathrm{c}}{O,t_{\smiley}} \equiv {\fn{P_\mathrm{c}}{O}} / {\fn{P_\varnothing}{O}}$,
which depends on the observing time $t_{\smiley}$.
Then \eq{eq:prior_vague} can be written as
\begin{equation}
\fn{P}{O|\smiley} = \fn{P_\varnothing}{O} \fn{W_\mathrm{c}}{O, t_{\smiley}} \fn{L_\mathrm{c}}{O|\smiley}.
\end{equation}

In the case that $O$ is the cosmological constant,
if one assumes that $\Lambda = 0$ is not unique,
then it seems reasonable to take $\fn{P_\varnothing}{\Lambda}$ to be constant over the anthropically interesting range of $\Lambda$,
since these values are very small compared with particle physics scales \cite{Weinberg:1987dv,Efstathiou:1995ne}.

$\fn{W_\mathrm{c}}{\Lambda, t_{\smiley}}$ is determined by the multiverse measure.
We consider three multiverse measures:
the pocket based measure \cite{Garriga:1998px,Vanchurin:1999iv,Garriga:2005av,Easther:2005wi},
the scale factor cutoff measure \cite{DeSimone:2008bq,Bousso:2008hz} and the causal patch measure \cite{Bousso:2006ev,Bousso:2006ge},
which correspond to counting the number of observers within a comoving volume,
a physical volume
and a Hubble volume, respectively.
In the case of the pocket based measure,
$\fn{W_\mathrm{c}}{\Lambda, t_{\smiley}} = 1$ by definition.
However, in the cases of the scale factor cutoff measure and the causal patch measure,
$\fn{W_\mathrm{c}}{\Lambda, t_{\smiley}}$ depends on the choice of $t_{\smiley}$.
Assuming that we are typical observers, and the preferred time is mostly determined by stellar evolution, we set the observing time $t_{\smiley}$ as the physical time $t_0$ (14 billion years)\footnote{The origin of time could also be chosen as the time when the density perturbation becomes of order unity,
but this would not make much difference to our results.

Lineweaver \& Egan \cite{Lineweaver:2007qh} estimated the age distribution of terrestrial planets, and argued that $t_{\smiley} = t_0$ may be a typical time for the terrestrial-type observers.}.

$\fn{L_\mathrm{c}}{\Lambda | \smiley}$ is determined by the number observers in a comoving volume,
and we will focus on it in the following sections.
In this section we will neglect its effect and set it as constant.

\begin{figure}[hbt]
\begin{center}
\includegraphics[height=0.3\textwidth]{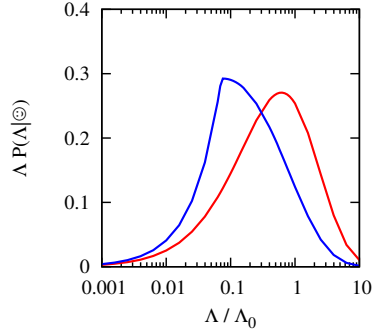}
\caption{\label{fig:prob_measure}
Normalized probability distribution $\Lambda \fn{P}{\Lambda|\smiley}$ for the {\color{red} scale factor cutoff} measure and the {\color{blue} causal patch} measure,
assuming both $\fn{P_\varnothing}{\Lambda}$ and $\fn{L_\mathrm{c}}{\Lambda|\smiley}$ are constant, and $t_{\smiley} = t_0$.
The pocket based measure is not shown here because it is very small.
We use $\Lambda \fn{P}{\Lambda|\smiley}$ to make the area inside the curve to be the probability.}
\end{center}
\end{figure}

\begin{table}[hbt]
\begin{center}
\begin{tabular}{|c||c|c|c|}
\hline $\typicality{\Lambda_0}$ & PB & SFC & CP \\[0.5ex]
\hline \hline $t_{\smiley} = t_0$ & $2 \times 10^{-120}$ & 0.55 & 0.14 \\[0.5ex]
\hline
\end{tabular}
\caption{\label{tab:prior_lambda}
Typicalities of the observed value of the cosmological constant, $\typicality{\Lambda_0}$, for the pocket based measure (PB), the scale factor cutoff measure (SFC), and the causal patch measure (CP), 
assuming both $\fn{P_\varnothing}{\Lambda}$ and $\fn{L_\mathrm{c}}{\Lambda|\smiley}$ are constant.
We assume $0 \lesssim \Lambda \lesssim 1$.}
\end{center}
\end{table}

\fig{fig:prob_measure} and \tab{tab:prior_lambda} show $\Lambda \fn{P}{\Lambda|\smiley}$ and the typicality \cite{Page:2006er} of $\Lambda_0$ for different multiverse measures, assuming $\fn{P_\varnothing}{\Lambda}$ and $\fn{L_\mathrm{c}}{\Lambda|\smiley}$ are constant so that only $\fn{W_\mathrm{c}}{\Lambda, t_{\smiley}}$ affects the shape of $\fn{P}{\Lambda|\smiley}$, where the typicality of an observable $O = O_0$ is defined as\footnote{In this paper we only consider the typicality within the range $O > 0$ and so normalize the probability as $\displaystyle \int_0^{\infty} \d{O} \fn{P}{O|\smiley} = 1$. The typicality using the whole range of $O$ is greater than that using $O > 0$ \cite{Pogosian:2006fx}.} \begin{equation} 
\fn{\mathcal{T}_+}{O_0} = 2 \times \min 
\biggl[ \int_0 ^{O_0 } \d{O} \, \fn{P}{O|\smiley}, 
\int_{O_0 } ^{\infty} \d{O} \, \fn{P}{O|\smiley}\biggr]\,.\label{eq:typicality} \end{equation} 
In the case of the pocket based measure, 
$\fn{W_\mathrm{c}}{\Lambda, t_{\smiley}} = 1$ so $\Lambda \fn{P}{\Lambda|\smiley}$ is low for $\Lambda \sim \Lambda_0$. 
In the cases of the scale factor cutoff measure and the causal patch measure, 
since a physical volume and a Hubble volume are smaller than a comoving volume in a large $\Lambda$, 
the scale factor cutoff and the causal patch measure suppress the region where $\Lambda$ dominates at $t = t_0$, 
which makes the typicality of $\Lambda_0$ for both measures high. 
See Appendix~\ref{app:prior} for analytic forms.

\begin{table}[hbt]
\centering
\begin{tabular}{|c||c|c|}
\hline $\typicality{Q_0}$ & $\fn{P_\mathrm{c}}{Q} = \textrm{constant}$ & $\fn{P_\mathrm{c}}{Q} \propto Q^{-1}$ \\[0.5ex]
\hline \hline - & $2 \times 10^{-5}$ & 0.63 \\[0.5ex]
\hline
\end{tabular}
\caption{\label{tab:prior_q}
Typicalities of the observed value of the primordial density perturbation amplitude, $\typicality{Q_0}$,
for different $\fn{P}{Q}$, assuming $\fn{L_\mathrm{c}}{Q|\smiley}$ is constant.
We assume $10^{-16} \lesssim Q \lesssim 1$.
}
\end{table}

In the case of the primordial density perturbation amplitude, 
since it does not affect the volume of the universe,
the multiverse ambiguity is independent of $t_{\smiley}$
so $\fn{W_\mathrm{c}}{Q, t_{\smiley}} = 1$.
If we also assume that $\fn{L_\mathrm{c}}{Q|\smiley}$ is constant,
then $\fn{P}{Q|\smiley}$ is determined only by $\fn{P_\varnothing}{Q}$.
Since $\fn{P_\varnothing}{Q}$ is unknown,
we consider two toy models for $\fn{P_\varnothing}{Q}$: flat in linear scale, i.e.\ $\fn{P_\varnothing}{Q} = \textrm{constant}$,
and flat in log scale, i.e.\ $\fn{P_\varnothing}{Q} \propto Q^{-1}$.
In the case $\fn{P_\varnothing}{Q} = \textrm{constant}$, the typicality is small,
which means that the prior distribution itself predicts much larger $Q$ than $Q_0$. 
On the other hand, 
since $Q_0 \sim 10^{-5}$ lays in the middle of a plausible range of $Q$ in the log scale,
$10^{-16} \lesssim Q \lesssim 1$,\footnote{We set the lower bound of $Q$ from $Q \gtrsim H_\mathrm{inflation} \gtrsim m_\mathrm{susy}$.} 
$\fn{P_\varnothing}{Q} \propto Q^{-1}$ gives relatively large $\typicality{Q_0}$.
See \tab{tab:prior_q}.

However, the actual probability of observable $O$ with taking into account of our existence is given by
\begin{equation}
\fn{P}{O|\smiley} = \fn{P_\varnothing}{O} \fn{W_\mathrm{c}}{O, t_{\smiley}} \fn{L_\mathrm{c}}{O|\smiley}.
\end{equation}
Therefore, our results can be changed significantly,
depending on the actual form of $\fn{L_\mathrm{c}}{O|\smiley}$.
Especially, even the pocket based measure and $\fn{P_\mathrm{c}}{Q} = \textrm{constant}$ may also be able to explain $\Lambda_0$ and $Q_0$ well,
by combining with the anthropic likelihood.

\section{Anthropic models using a single mass constraint}
\label{sec:fin}

\subsection{Weinberg's anthropic model: $M \geq M_*$ at $t \to \infty$}
\label{sec:fin_classic}

Weinberg et al.\ \cite{Martel:1997vi} and Vilenkin et al.\ \cite{Pogosian:2006fx} model the number of observers as proportional to the total mass in gravitationally collapsed objects with mass greater than a certain threshold, $M \geq M_*$, at late times, $t \to \infty$.
There are several motivations for this model:
\begin{enumerate}
\item
Uncollapsed mass is not expected to give rise to observers.
\item
The total mass of gravitationally collapsed objects is one of the easiest quantities to calculate.
\item
If the collapsed object is too small, then there may be no chance for the evolution of complex life, for example, due to lack of metals.
\item
Once a collapsed object is large enough to be habitable, the number of observers may plausibly be proportional to the number of baryons in the object and hence proportional to the total mass.
\item
Once the mass of an object overcomes $M_*$, it may become habitable irrespective of when it formed, and so the collapsed objects with $M \geq M_*$ at $t \to \infty$ may include all habitable collapsed objects.
\end{enumerate}

Weinberg's model includes our object, the Milky Way,
but it also includes supermassive objects.
However, supermassive objects may not be very habitable, for example, due to the strong interactions between galaxies in superclusters.
If this is true, including these supermassive objects gives a bias to small cosmological constant, since a small cosmological constant lets matter collapse more easily.
Thus, Weinberg's model may give a misleadingly good result.

The choice of $M_*$ in Weinberg's model is another difficulty.
If $M_*$ is taken to be the mass of the Milky Way, as is usually done, then the typical mass in $M \geq M_*$ is greater than that of the Milky Way and we are not typical.
This anomaly can be reduced by choosing smaller $M_*$, but reducing $M_*$ reduces the typicality, and there is no obvious smaller choice of $M_*$.
\subsection{$M = M_*$ at $t = t_*$}
\label{sec:fin_fix}
This model assumes that there exist anthropically preferred mass and time scales $M_*$ and $t_*$,
so that the number of observers is proportional to the fraction of gravitationally collapsed objects with $M = M_*$ at $t = t_*$.
\footnote{Graesser \& Salem \cite{Graesser:2006ft} also used $M = M_*$ but kept $t \to \infty$.}

In order to choose $M_*$ and $t_*$ we will use the assumption that we are typical observers,
although one must be careful not to introduce bias by considering features due to our value of the cosmological constant as opposed to features affecting the formation of observers.

By assuming we are typical observers, it seems obvious to choose $M_*$ as the mass of the Milky Way and $t_*$ as $t_0$ (14 billion years).
However, the Press-Schechter formalism \cite{Press:1973iz}, which we use to calculate the fraction of collapsed objects, identifies the Local Group, not the Milky Way, at $t = t_0$.
This is because the formalism can only identify objects of the mass of the Milky Way which are isolated within at least $1.9\Mpc$ at $t = t_0$, but Andromeda and other members of the Local Group are now within this range.
Thus, the Press-Schechter formalism seems to require us to choose $M_*$ as the mass of the Local Group, but as we do not seem to have any plausible anthropic justification to use the Local Group as our object
\footnote{Other members of a group of galaxies may perturb merging objects away from direct hit trajectories which may be anthropically beneficial.}, this would not be consistent either.

However, it is not only the present time which affects our existence.
For example, the state of the galaxy before the formation of the solar system may be essential by influencing the star formation rate or metal abundance.
Also, galaxies may need to be isolated up to a certain time, in order to prevent harmful interactions.
Thus, we may have anthropic motivation to choose $t_*$ earlier than the formation of the solar system,
or even earlier than the formation of the Local Group.
If we set $t_*$ earlier than the formation of the Local Group,
the technical problem with using the Press-Schechter formalism disappears,
since in this case we can identify the Milky Way as an isolated collapsed object. 
Thus, we set $t_*$ as a time earlier than the formation of the Local Group and $M_*$ as the mass of the Milky Way at that time.

Here, as a definite example, we take $t_*$ as 6 billion years.
Also, Refs.~\cite{Hammer:2007ki,Burstein:2004pn} suggest that the Milky Way may not have had any major interaction or a significant amount of minor mergers over the last 10 billion years, so we may approximate the mass of the Milky Way at 6 billion years as similar to its current mass.
Thus, we set $M_*$ as the mass of the Milky Way ($M_\mathrm{MW}$).

\subsection{Results}
\label{sec:fin_res}

\begin{figure}[hbt]
\centering
\includegraphics[height=0.3\textwidth]{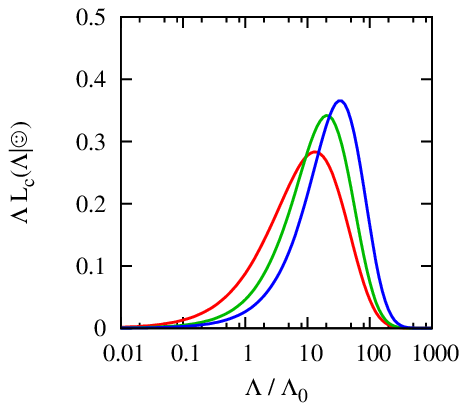} \hspace{20pt}
\includegraphics[height=0.3\textwidth]{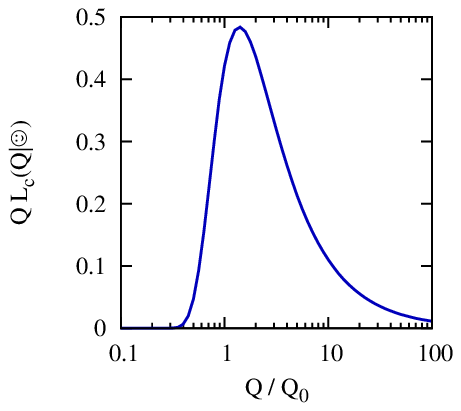}
\caption{Anthropic likelihoods for the anthropic models using a single mass constraint. 
Left: anthropic likelihoods for the cosmological constant $\fn{L_\mathrm{c}}{\Lambda|\smiley}$.
Anthropic model: {\color{red} $M \geq M_\mathrm{MW}$ at $t \to \infty$}, {\color[rgb]{0,0.5,0} $M \geq M_\mathrm{MW}$ at $t = 6\Gyr$} and {\color{blue} $M = M_\mathrm{MW}$ at $t = 6\Gyr$}.
Right: anthropic likelihoods for the primordial density perturbation amplitude $\fn{L_\mathrm{c}}{Q|\smiley}$, for $\Lambda = \Lambda_0$.
$\fn{L_\mathrm{c}}{Q|\smiley}$ for anthropic models with $M \geq M_\mathrm{MW}$ is not shown because it is very small.
}\label{fig:final}
\end{figure}

In addition to Weinberg's model, $M \geq M_\mathrm{MW}$ at $t \to \infty$, and the model $M = M_\mathrm{MW}$ at $t = 6\Gyr$,
we consider the model $M \geq M_\mathrm{MW}$ at $t = 6\Gyr$
to see how the mass and time conditions independently affect the likelihood.
We also consider the anthropic likelihood for the primordial density perturbation amplitude, 
assuming $\Lambda = \Lambda_0$.
\fig{fig:final} shows both $\fn{L_\mathrm{c}}{\Lambda|\smiley}$ and $\fn{L_\mathrm{c}}{Q|\smiley}$ for each model
(see Appendix~\ref{app:final} for the analytic forms).

\begin{figure}[hbt]
\centering
\includegraphics[height=0.3\textwidth]{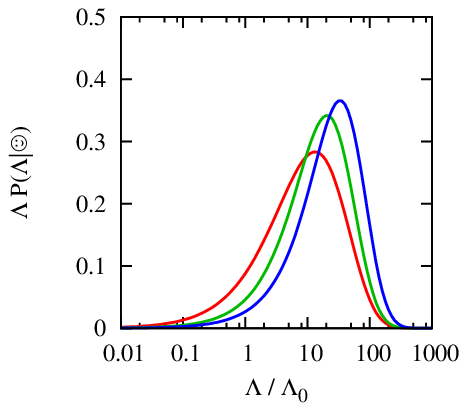}
\includegraphics[height=0.3\textwidth]{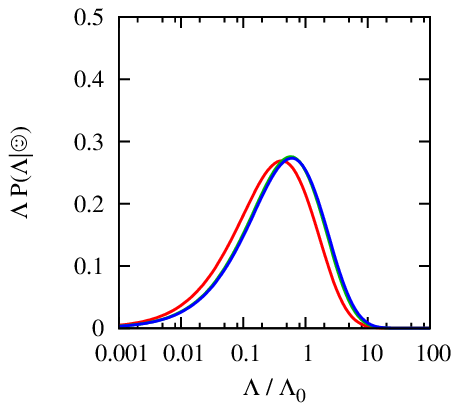}
\includegraphics[height=0.3\textwidth]{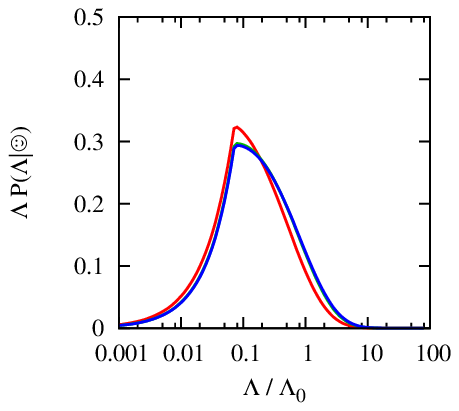}
\caption{Probability of an observer observing $\Lambda$, $\fn{P}{\Lambda|\smiley}$.
Left: pocket based measure; Middle: scale factor cutoff measure with $t_\mathrm{obs} = t_0$;
Right: causal patch measure with $t_\mathrm{obs} = t_0$.
Anthropic model: {\color{red} $M \geq M_\mathrm{MW}$ at $t \to \infty$}, {\color[rgb]{0,0.5,0} $M \geq M_\mathrm{MW}$ at $t = 6\Gyr$} and {\color{blue} $M = M_\mathrm{MW}$ at $t = 6\Gyr$}.}\label{fig:final_lambda}
\end{figure}

\begin{table}[hbt]
\centering
\begin{tabular}{|r @{~at~} l ||c|c|c|}
\hline \multicolumn{2}{|c||}{\multirow{2}{*}{$\typicality{\Lambda_0}$}} & \multirow{2}{*}{PB} & SCF & CP \\
\multicolumn{2}{|c||}{} & & $t_{\smiley} = t_0$ & $t_{\smiley} = t_0$ \\[0.5ex]
\hline \hline $M \geq M_\mathrm{MW}$ & $t \to \infty$ & 0.22 & 0.36 & 0.11 \\[0.5ex]
\hline $M \geq M_\mathrm{MW}$ & $t = 6\Gyr$ & 0.086 & 0.49 & 0.16 \\[0.5ex]
\hline $M = M_\mathrm{MW}$ & $t = 6\Gyr$ & 0.049 & 0.52 & 0.17 \\[0.5ex]
\hline
\end{tabular}
\caption{\label{tab:final_lambda}
$\typicality{\Lambda_0}$ for the anthropic models using a single mass constraint for the pocket based measure (PB), the scale factor cutoff measure (SFC), and the causal patch measure (CP).}
\end{table}

\begin{table}[hbt]
\centering
\begin{tabular}{|r @{~at~} l ||c|c|}
\hline \multicolumn{2}{|c||}{$\typicality{Q_0}$} & $\fn{P_\mathrm{c}}{Q} = \mathrm{constant}$ & $\fn{P_\mathrm{c}}{Q} \propto Q^{-1}$ \\[0.5ex]
\hline \hline $M \geq M_\mathrm{MW}$ & $t \to \infty$ & $7 \times 10^{-7}$ & $8 \times 10^{-3}$ \\[0.5ex]
\hline $M \geq M_\mathrm{MW}$ & $t = 6\Gyr$ & $1 \times 10^{-7}$ & $2 \times 10^{-3}$ \\[0.5ex]
\hline $M = M_\mathrm{MW}$ & $t = 6\Gyr$ & 0.33 & 0.76 \\[0.5ex]
\hline
\end{tabular}
\caption{\label{tab:final_Q}
$\typicality{Q_0}$ for the anthropic models using a single mass constraint, for $\Lambda = \Lambda_0$.
We assume $10^{-16} \lesssim Q \lesssim 1$.}
\end{table}

\fig{fig:final_lambda} and \twotab{tab:final_lambda}{tab:final_Q} summarize the typicalities in the different anthropic models.
In the case of the pocket based measure which Weinberg et al.\ \cite{Martel:1997vi} implicitly used,
$\typicality{\Lambda_0}$ decreases by a factor of two as $t_*$ changes from infinity to $6\Gyr$, 
and decreases by a further factor of two as the constraint changes from $M \geq M_\mathrm{MW}$ to $M = M_\mathrm{MW}$.
This illustrates how Weinberg's model may overestimate the typicality.
On the other hand, in the cases of the scale factor cutoff and causal patch measures,
the prior distribution from the measure already suppresses the region where the difference between the anthropic likelihoods from the different anthropic models is significant. 
Therefore, in these cases,
all three anthropic models provide typicalities similar to the one only assuming $t_{\smiley} = t_0$.

In the case of the primordial density perturbation amplitude,
anthropic models with the mass constraint $M \geq M_\mathrm{MW}$ include supermassive objects,
which always prefer large $Q$.
Therefore, anthropic models with $M \geq M_\mathrm{MW}$ give a low typicality of $Q_0$,
i.e.\ beyond $3\sigma$.
So Weinberg's model may require the extra anthropic bound of $Q$ suggested by Tegmark \& Rees \cite{Tegmark:1997in}, 
$10^{-1} Q_0 \lesssim Q \lesssim 10 Q_0$,
which ensures sufficient cooling of galaxies and the stable orbits of planets.
On the other hand,
our model $M = M_\mathrm{MW}$ at $t = 6\Gyr$ gives a high typicality of $Q_0$, i.e.\ within $1\sigma$, without any additional assumption.
Note that the choice of the prior distribution for $Q$ does not provide any qualitative difference.

\subsection{Degeneracies between $\Lambda$ and $Q$}\label{sec:fin_Q}

In principle, one should analyze the entire space of physical parameters to determine the anthropic likelihood of the cosmological constant.
A first step toward this direction is to examine the two dimensional parameter space of $\Lambda$ and $Q$.
In this two dimensional parameter space, larger primordial density perturbation amplitude can cancel the effect of the cosmological constant \cite{Tegmark:1997in,Banks:2003es, Graesser:2004ng,Garriga:2005ee},
leading to degeneracies in the parameter space.

\begin{figure}[hbt]
\centering
\includegraphics[height=0.3\textwidth]{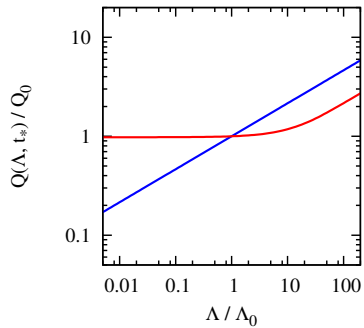}
\caption{ \label{fig:Q_lamb}
$\fn{Q}{\Lambda,t_*}$ for which the population of galaxies and clusters at $t = t_*$ is independent of $\Lambda$, for {\color{red} $t_* = 6\Gyr$} and {\color{blue} $t_* \to \infty$}.
}
\end{figure}

For simplicity, we choose a slice from the $(\Lambda, Q)$ space which maximizes the degeneracy.
We set $Q = \fn{Q}{\Lambda,t_*}$ so that the value of the matter power spectrum on the scales of galaxies at $t = t_*$ is independent of the cosmological constant.
\fig{fig:Q_lamb} shows $\fn{Q}{\Lambda,t_*}$, and it has the large $\Lambda$ behavior
\begin{equation}
\fn{Q}{\Lambda,t_*} \propto \Lambda^{\frac{1}{3}}
\qquad \textrm{for } \Lambda / \Lambda_0 \gg \fn{f}{{t_*}/{t_0}}\,,
\end{equation}
where $\fn{f}{6\Gyr / t_0} \simeq 10$, and the late time behavior
\begin{equation}
\fn{Q}{\Lambda,\infty} = Q_0 \left(\frac{\Lambda}{\Lambda_0}\right)^{\frac{1}{3}} .
\end{equation}
See Appendix~\ref{app:sigma_qfunc} for the exact form of $\fn{Q}{\Lambda,t_*}$.

On the slice $Q = \fn{Q}{\Lambda,t_*}$, the probability of observing $\Lambda = \Lambda'$ is
\footnote{We define $\displaystyle \fn{P_\mathrm{c}}{\fn{Q}{\Lambda,t_*}} \equiv \int \d{\Lambda'} \fn{P_\mathrm{c}}{\Lambda=\Lambda',Q=\fn{Q}{\Lambda',t_*}}$.}
\begin{equation}
\fn{P}{\Lambda=\Lambda'|\smiley,\fn{Q}{\Lambda,t_*}} = 
\frac{ \fn{P}{\Lambda=\Lambda'|\fn{Q}{\Lambda,t_*}} \fn{P}{\smiley|\Lambda=\Lambda',\fn{Q}{\Lambda,t_*}} }{ \fn{P}{\smiley|\fn{Q}{\Lambda,t_*}} } .
\end{equation}
For anthropic models using a single time $t = t_*$ and $Q = \fn{Q}{\Lambda,t_*}$,
the population of galaxies and clusters at $t_*$ is independent of the cosmological constant,
and so $\fn{P}{\smiley|\Lambda=\Lambda',\fn{Q}{\Lambda,t_*}}$ is independent of $\Lambda'$.
Therefore, the anthropic likelihood becomes
\begin{equation}
\fn{L_\mathrm{c}}{\Lambda=\Lambda'|\smiley,\fn{Q}{\Lambda,t_*}} \equiv \left. \frac{ \fn{P}{\smiley|\Lambda=\Lambda',\fn{Q}{\Lambda,t_*}} }{ \fn{P}{\smiley|\fn{Q}{\Lambda,t_*}} } \right|_\mathrm{c} = 1 ,
\end{equation}
and the probability of observing $\Lambda = \Lambda'$ reduces to the modified prior,
\begin{align}
\fn{P}{\Lambda=\Lambda'|\smiley,\fn{Q}{\Lambda,t_*}} & = \fn{P_\mathrm{c}}{\Lambda=\Lambda'|\fn{Q}{\Lambda,t_*}} \\
& = \frac{ \fn{P_\mathrm{c}}{\Lambda=\Lambda'} \fn{P_\mathrm{c}}{\fn{Q}{\Lambda,t_*}|\Lambda=\Lambda'} }{ \fn{P_\mathrm{c}}{\fn{Q}{\Lambda,t_*}} } .
\end{align}
Therefore,
\begin{equation}
\frac{ \fn{P}{\Lambda=\Lambda'|\smiley,\fn{Q}{\Lambda,t_*}} }{ \fn{P_\mathrm{c}}{\Lambda=\Lambda'} }
\propto \fn{P_\mathrm{c}}{Q=\fn{Q}{\Lambda',t_*}}
\end{equation}
depends on the prior distribution of $Q$.

For example,
if $\fn{P_\mathrm{c}}{Q} = \textrm{constant}$,
\begin{equation}
\frac{ \fn{P}{\Lambda=\Lambda'|\smiley,\fn{Q}{\Lambda,t_*}} }{ \fn{P_\mathrm{c}}{\Lambda=\Lambda'} } = \textrm{constant} ,
\end{equation}
or if $\fn{P_\mathrm{c}}{Q} \propto Q^{-1}$,
\begin{equation}
\frac{ \fn{P}{\Lambda=\Lambda'|\smiley,\fn{Q}{\Lambda,t_*}} }{ \fn{P_\mathrm{c}}{\Lambda=\Lambda'} } \propto {\Lambda'}^{-\frac{1}{3}}
\qquad \textrm{for } \Lambda' / \Lambda_0 \gg \fn{f}{{t_*}/{t_0}} 
\end{equation}

On the other hand,
if we apply Tegmark \& Rees' anthropic bound of $Q$ \cite{Tegmark:1997in},
$10^{-1} Q_0 \lesssim Q \lesssim 10 Q_0$,
$\Lambda$ can be also constrained on the slice $Q = \fn{Q}{\Lambda,t_*}$ as $10^{-3} \Lambda_0 \lesssim \Lambda \lesssim 10^3 \Lambda_0$,
which effectively breaks the degeneracy between $\Lambda$ and $Q$.
See \tab{tab:Q_lamb}.

\begin{table}[hbt]
\centering
\begin{tabular}{|r @{~$\lesssim$~} c @{~$\lesssim$~} l ||c|c|}
\hline \multicolumn{3}{|c||}{\multirow{2}{*}{$\typicality{\Lambda_0}$}} & $Q = \fn{Q}{\Lambda, t_\mathrm{f}}$ & $Q = \fn{Q}{\Lambda, t_\mathrm{f}}$ \\[0.5ex]
\multicolumn{3}{|c||}{ } & $\fn{P_\mathrm{c}}{Q} = \textrm{constant}$ & $\fn{P_\mathrm{c}}{Q} \propto Q^{-1}$ \\[0.5ex]
\hline \hline 0 & $\Lambda$ & 1 & $2 \times 10^{-120}$ & $2 \times 10^{-80}$ \\[0.5ex]
\hline $10^{-16}$ & $Q$ & 1 & $2 \times 10^{-15}$ & $2 \times 10^{-10}$ \\[0.5ex]
\hline $10^{-6}$ & $Q$ & $10^{-4}$ & $2 \times 10^{-3}$ & $2 \times 10^{-2}$ \\[0.5ex]
\hline
\end{tabular}
\caption{\label{tab:Q_lamb}
$\typicality{\Lambda_0}$ for different boundaries of $\Lambda$ and $Q$,
assuming $Q = \fn{Q}{\Lambda, t_\mathrm{f}}$ for which the population of galaxies and clusters at $t = t_\mathrm{f}$ is independent of $\Lambda$.
We assume flat prior/pocket based measure.
$10^{-16} \lesssim Q \lesssim 1$ and $10^{-6} \lesssim Q \lesssim 10^{-4}$ come from $Q \gtrsim H_\mathrm{inflation} \gtrsim m_\mathrm{susy}$ and Tegmark \& Rees \cite{Tegmark:1997in}, respectively.
Note that the typicality does not depend on the mass constraint.}
\end{table}

\section{Anthropic models using the mass history}
\label{sec:his}

\subsection{Motivation}
\label{sec:his_mot}
The evolution of life and creation of observers depends on many complex factors.
For example, early accretion may determine the population of early stars in galaxies, which determines the element abundance of later stellar gas which is crucial to the formation of complex life.
Mergers or collisions of galaxies may damage or destroy habitable environments within galaxies, for example, by disturbing peaceful stellar orbits, triggering star formation and supernovae, or activating galactic nuclei.
These features, which may be beneficial or harmful for the formation of observers, cannot be taken into account by considering just the mass of a gravitationally collapsed object at a single time.
As a first step towards taking into account these complex factors, we will consider the mass history of the gravitationally collapsed object.

\subsection{Calculational technique: extended Press-Schechter}
\label{sec:his_meth}
To calculate the anthropic likelihood taking into account the mass history,
we use the extended Press-Schechter formalism \cite{Bond:1990iw,Bower:1991kf,Lacey:1993iv}.
The extended Press-Schechter formalism computes the mass fraction of collapsed objects with $M = M_\mathrm{f}$ at a certain time $t = t_\mathrm{f}$,
which were formed from objects with $M = M_\mathrm{i}$ at an earlier time $t = t_\mathrm{i}$.
This formalism limits us to taking into account only two points in the mass history to determine the anthropic likelihood.

\subsection{Results}
\label{sec:his_res}

\begin{figure}[hbt]
\centering
\includegraphics[height=0.3\textwidth]{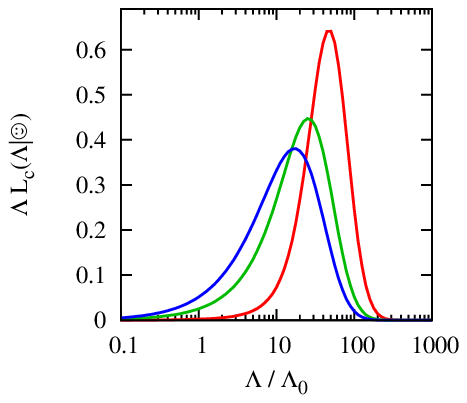}
\includegraphics[height=0.3\textwidth]{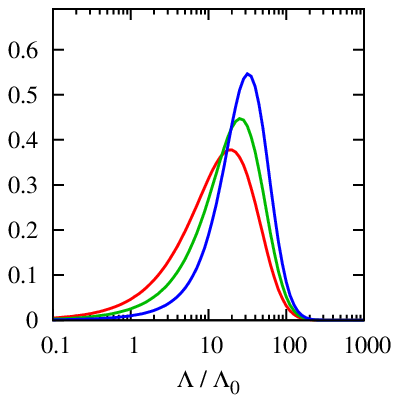}
\includegraphics[height=0.3\textwidth]{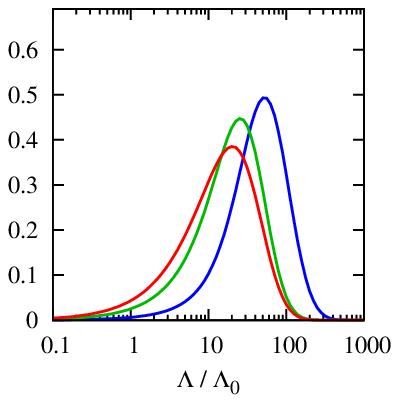} \\[10pt]
\includegraphics[height=0.3\textwidth]{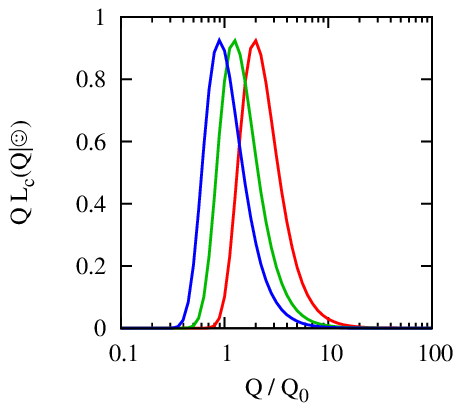}
\includegraphics[height=0.3\textwidth]{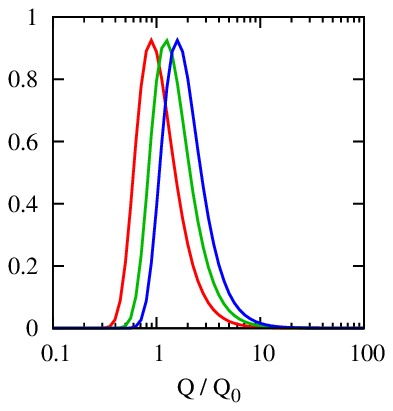}
\includegraphics[height=0.3\textwidth]{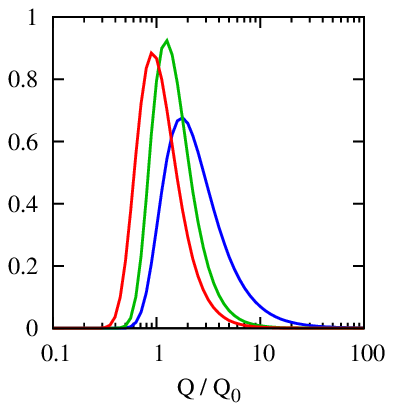} 
\caption{\label{fig:his_L}
$\fn{L_\mathrm{c}}{\Lambda|\smiley}$ (top) and $\fn{L_\mathrm{c}}{Q|\smiley}$ (bottom) for the anthropic models using the mass history with $M_\mathrm{f} = M_\mathrm{MW}$ at $t_\mathrm{f} = 6\Gyr$.
Left: $M_\mathrm{i} = 0.8 M_\mathrm{MW}$ at $t_\mathrm{i} = $ {\color{red} $3\Gyr$}, {\color[rgb]{0,0.5,0} $4\Gyr$} and {\color{blue} $5\Gyr$} .
Middle: $M_\mathrm{i} = $ {\color{red} $0.2 M_\mathrm{MW}$}, {\color[rgb]{0,0.5,0} $0.8 M_\mathrm{MW}$} and {\color{blue} $0.9 M_\mathrm{MW}$} at $t_\mathrm{i} = 4\Gyr$.
Right: $M_\mathrm{i}$ {\color{red} $\leq$}, {\color[rgb]{0,0.5,0} $=$} and {\color{blue} $\geq$} $0.8 M_\mathrm{MW}$ at $t_\mathrm{i} = 4\Gyr$.
}
\end{figure}

In \sect{sec:fin},
we used the anthropic model with a single mass constraint, $M = M_\mathrm{MW}$ at $t = 6\Gyr$.
We calculated the corresponding anthropic likelihoods and typicalities of the cosmological constant and the primordial density perturbation amplitude.
Here, in addition to the final mass constraint $M_\mathrm{f} = M_\mathrm{MW}$ at $t_\mathrm{f} = 6\Gyr$,
we consider three types of initial mass constraint $M_\mathrm{i}$ at an earlier time $t_\mathrm{i}$: $M_\mathrm{i} \geq M_*$, $M_\mathrm{i} = M_*$ or $M_\mathrm{i} \leq M_*$,
where $M_*$ is a certain mass scale.
As a central example, we set $M_* = 0.8 M_\mathrm{MW}$ and $t_\mathrm{i} = 4\Gyr$.

\begin{figure}[hbt]
\centering
\includegraphics[height=0.274\textwidth]{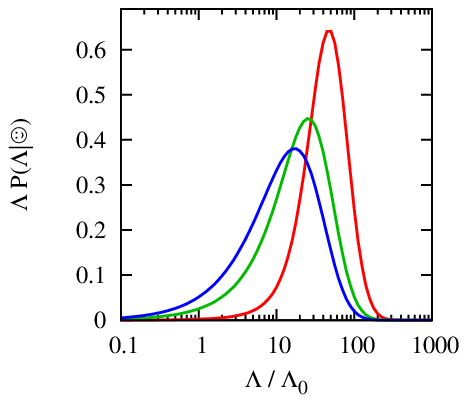}
\includegraphics[height=0.274\textwidth]{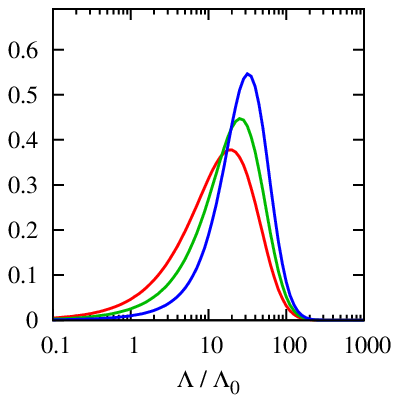} 
\includegraphics[height=0.274\textwidth]{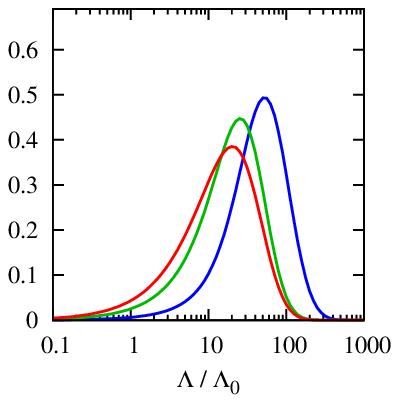} \\[2ex]
\includegraphics[height=0.274\textwidth]{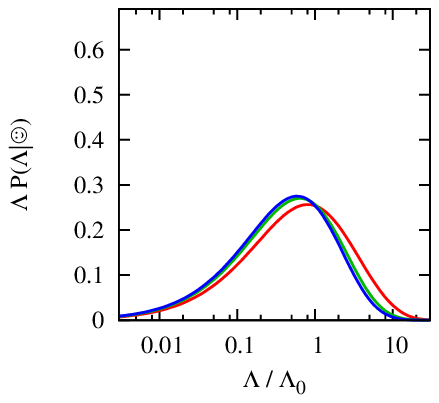} 
\includegraphics[height=0.274\textwidth]{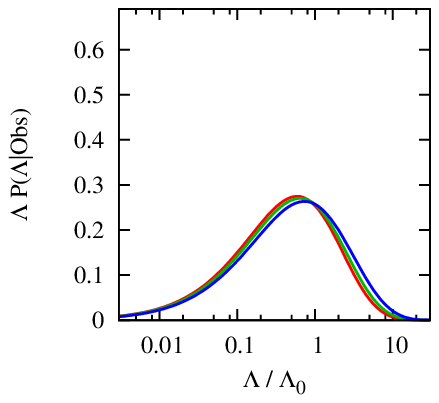} 
\includegraphics[height=0.274\textwidth]{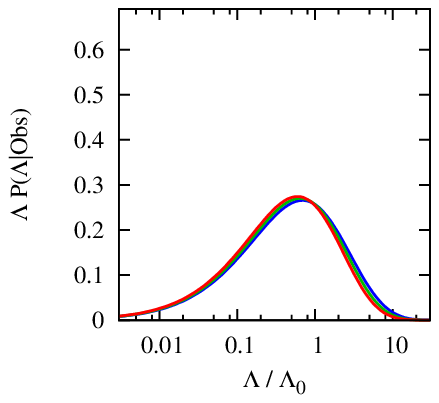} \\[2ex]
\includegraphics[height=0.3\textwidth]{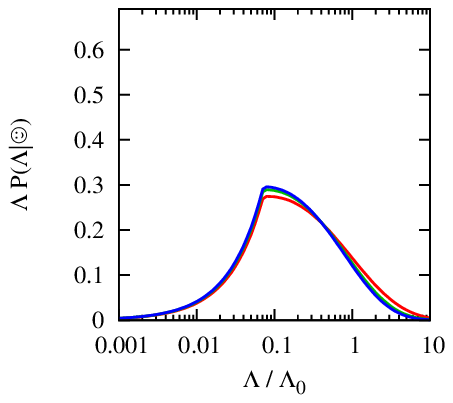}
\includegraphics[height=0.3\textwidth]{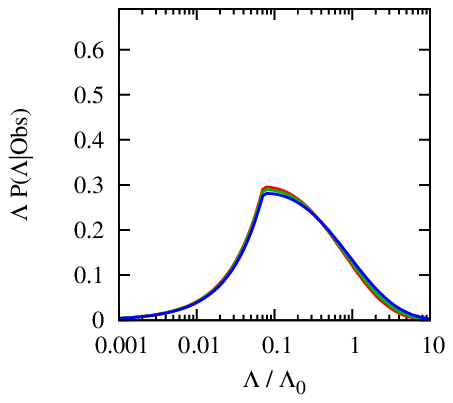}
\includegraphics[height=0.3\textwidth]{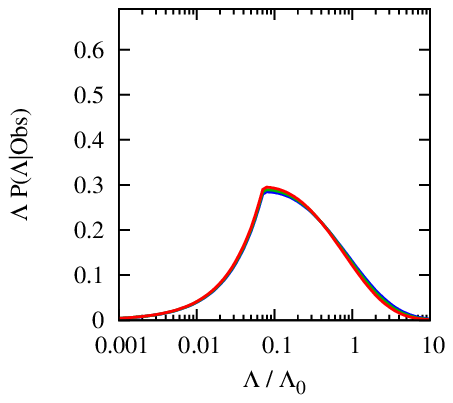}\\
\caption{\label{fig:his_measure}
$\fn{P}{\Lambda|\smiley}$ using the mass history with different multiverse measure.
Left to right: same to \fig{fig:his_L}.
Top: the pocket based measure; Middle: the scale factor cutoff measure; Bottom: the causal patch measure.}
\end{figure}

\fig{fig:his_L} shows the dependence of the anthropic likelihoods of $\Lambda$ and $Q$ on the choice of $t_\mathrm{i}$, $M_*$ and the mass constraint.
See Appendix~\ref{app:history} for the analytic forms.
Since matter collapses at later times if $\Lambda$ and $Q$ is smaller,
larger $t_\mathrm{i}$ and smaller $M_\mathrm{i}$ shift both $\fn{L_\mathrm{c}}{\Lambda|\smiley}$ and $\fn{L_\mathrm{c}}{Q|\smiley}$
toward smaller $\Lambda$ and $Q$.
However, in the cases of the scale factor cutoff and the causal patch measures,
the prior distribution suppresses the region where the change in the anthropic likelihood occurs,
and $\fn{P}{\Lambda|\smiley}$ remains similar regardless of the change of constraint (see \fig{fig:his_measure}).

\begin{figure}[p]
\centering
\includegraphics[height=0.274\textwidth]{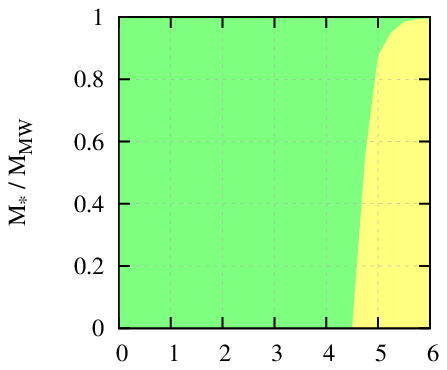}
\includegraphics[height=0.274\textwidth]{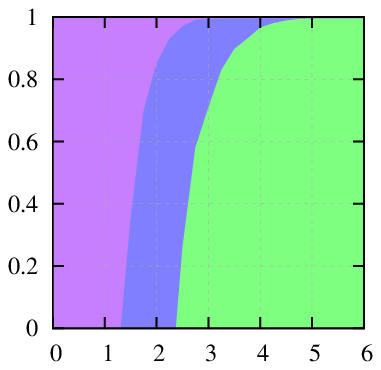}\\[2ex]
\includegraphics[height=0.274\textwidth]{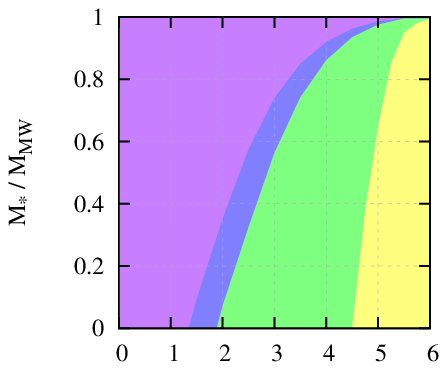} 
\includegraphics[height=0.274\textwidth]{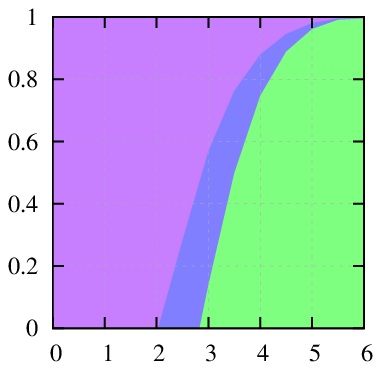}\\[2ex]
\includegraphics[height=0.3\textwidth]{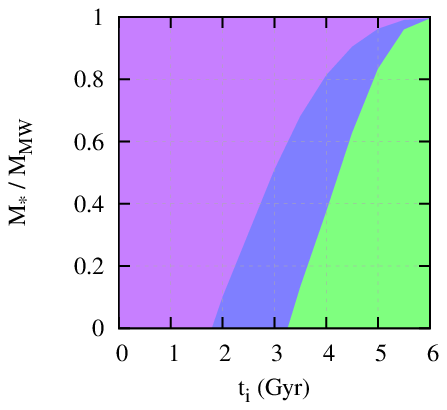}
\includegraphics[height=0.3\textwidth]{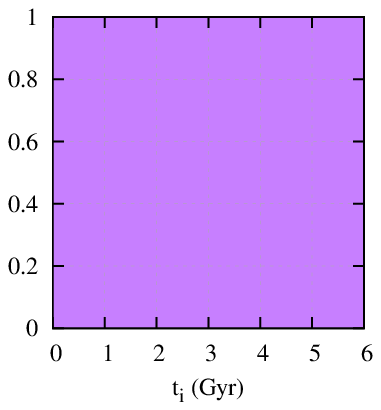}\\
\caption{\label{fig:his_cont_degeneracy}
Contour diagrams of $\fn{\mathcal{T}_+}{\Lambda_0}$
using the mass history with the usual flat prior/pocket based measure and $M_\mathrm{f} = M_\mathrm{MW}$ at $t_\mathrm{f} = 6\Gyr$.
Left: $Q = Q_0$; Right: $Q = \fn{Q}{\Lambda,t_\mathrm{f}}$,
assuming $\fn{P_\mathrm{c}}{Q} = \textrm{constant}$,
see \sect{sec:fin_Q}.
Top: $M_\mathrm{i} \leq M_*$; Middle: $M_\mathrm{i} = M_*$; Bottom: $M_\mathrm{i} \geq M_*$.
Typicality: {\color{violet} 0--0.01}, {\color{blue} 0.01--0.03}, {\color[rgb]{0,0.5,0} 0.03--0.1}, {\color[rgb]{0.4,0.4,0} 0.1--0.3}, {\color{red} 0.3--1}.}
\end{figure}

In order to understand how the typicality changes by mass and time constraints,
we plot in \figs{fig:his_cont_degeneracy}{fig:his_cont_lambda} the contour diagrams of the typicalities for the three types of constraint as a function of $t_\mathrm{i}$ and $M_*$.

\begin{figure}[p]
\centering
\includegraphics[height=0.274\textwidth]{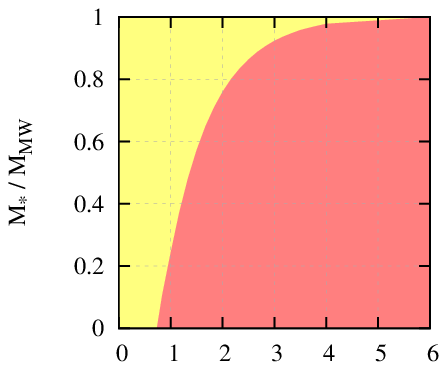} 
\includegraphics[height=0.274\textwidth]{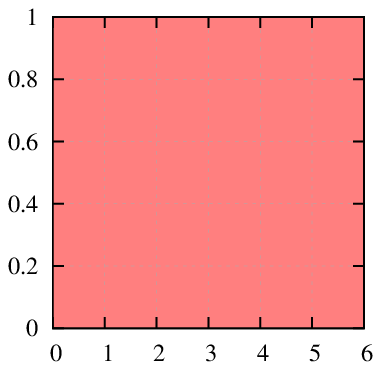}\\[2ex]
\includegraphics[height=0.274\textwidth]{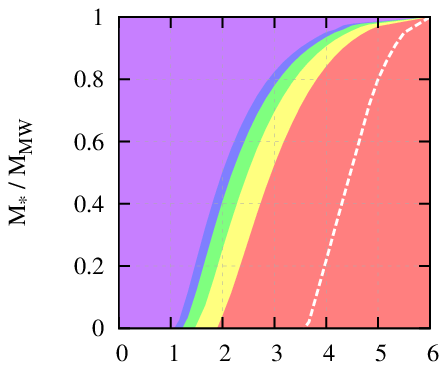} 
\includegraphics[height=0.274\textwidth]{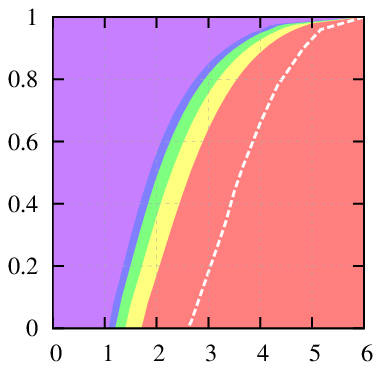}\\[2ex]
\includegraphics[height=0.3\textwidth]{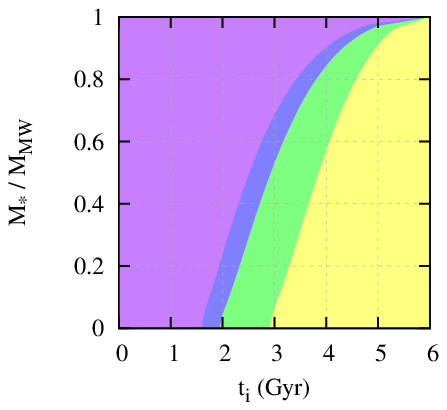} 
\includegraphics[height=0.3\textwidth]{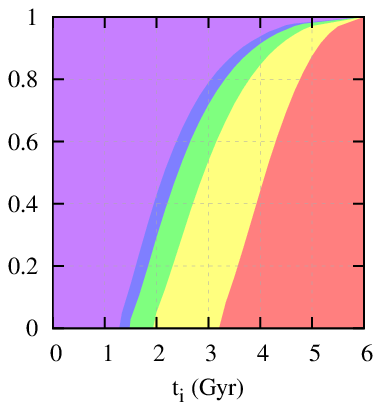} \\
\caption{\label{fig:his_cont_Q}
Contour diagrams of $\typicality{Q_0}$
using the mass history with $\Lambda = \Lambda_0$ and $M_\mathrm{f} = M_\mathrm{MW}$ at $t_\mathrm{f} = 6\Gyr$.
Left: $\fn{P_\mathrm{c}}{Q} = \textrm{constant}$; Right: $\fn{P_\mathrm{c}}{Q} \propto Q^{-1}$.
Top: $M_\mathrm{i} \leq M_*$; Middle: $M_\mathrm{i} = M_*$; Bottom: $M_\mathrm{i} \geq M_*$.
Typicality: {\color{violet} 0--0.01}, {\color{blue} 0.01--0.03}, {\color[rgb]{0,0.5,0} 0.03--0.1}, {\color[rgb]{0.4,0.4,0} 0.1--0.3}, {\color{red} 0.3--1}.
White dash: $\typicality{Q_0} = 1$.}
\end{figure}

\begin{figure}[p]
\centering
\includegraphics[height=0.274\textwidth]{eps/mw_6gyr_upper_pb}
\includegraphics[height=0.274\textwidth]{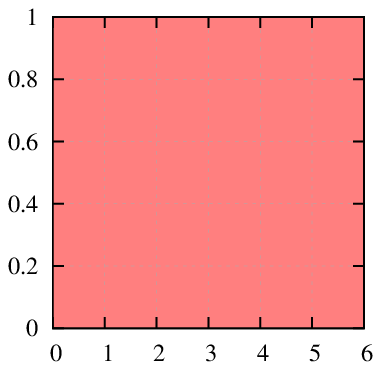} 
\includegraphics[height=0.274\textwidth]{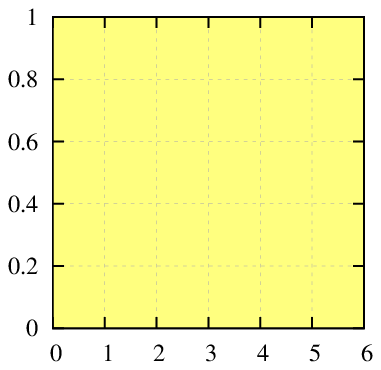} \\[2ex]
\includegraphics[height=0.274\textwidth]{eps/mw_6gyr_point_pb} 
\includegraphics[height=0.274\textwidth]{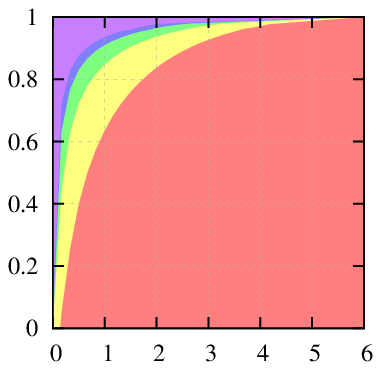} 
\includegraphics[height=0.274\textwidth]{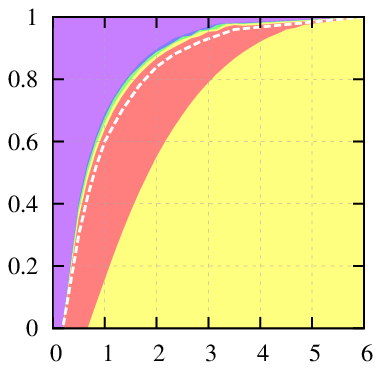} \\[2ex]
\includegraphics[height=0.3\textwidth]{eps/mw_6gyr_lower_pb}
\includegraphics[height=0.3\textwidth]{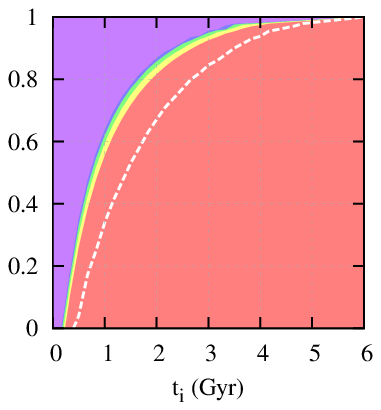}
\includegraphics[height=0.3\textwidth]{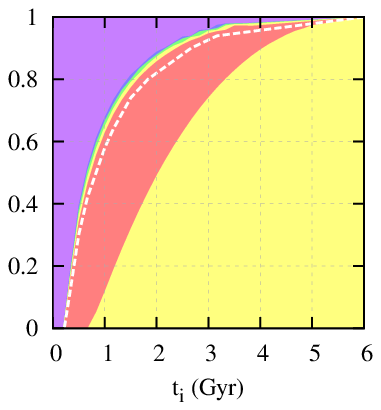}\\
\caption{\label{fig:his_cont_lambda}
Contour diagrams of $\typicality{\Lambda_0}$
using the mass history with $Q = Q_0$ and $M_\mathrm{f} = M_\mathrm{MW}$ at $t_\mathrm{f} = 6\Gyr$.
Left: pocket based measure; Middle: scale factor cutoff measure with $t_\mathrm{obs} = t_0$; Right: causal patch measure with $t_\mathrm{obs} = t_0$.
Top: $M_\mathrm{i} \leq M_*$; Middle: $M_\mathrm{i} = M_*$; Bottom: $M_\mathrm{i} \geq M_*$.
Typicality: {\color{violet} 0--0.01}, {\color{blue} 0.01--0.03}, {\color[rgb]{0,0.5,0} 0.03--0.1}, {\color[rgb]{0.4,0.4,0} 0.1--0.3}, {\color{red} 0.3--1}.
White dash: $\typicality{\Lambda_0} = 1$.}
\end{figure}

In \fig{fig:his_cont_degeneracy}, we start from the standard flat prior/pocket based measure,
and compare the cases $Q = Q_0$ and $Q = \fn{Q}{\Lambda,t_\mathrm{f}}$,
which makes the population of galaxies and clusters at $t = t_\mathrm{f}$ to be independent of $\Lambda$.
In the case of $Q = \fn{Q}{\Lambda, t_\mathrm{f}}$, we take $\fn{P_\mathrm{c}}{Q} = \textrm{constant}$,
which gives the greatest difference to the case of $Q = Q_0$.
Note that even in the case of $Q = \fn{Q}{\Lambda,t_\mathrm{f}}$
the value of the matter power spectrum on the scale of galaxies at the earlier time $t = t_\mathrm{i}$ depends on $\Lambda$.
Therefore, the degeneracy between $\Lambda$ and $Q$, 
discussed in \sect{sec:fin_Q} and which afflicts models using only a single mass constraint, 
is broken for models using the mass history.
However, the case $M_\mathrm{i} \geq M_*$ allows large $\Lambda$ and $Q$ and so the degeneracy is effectively unbroken.
For any history, the maximum value of typicality is $\typicality{\Lambda_0} \simeq 0.1$, i.e.\ about $1.5 \sigma$.

In \fig{fig:his_cont_Q}, we calculate the typicality of $Q_0$,
by considering the prior distributions $\fn{P_\mathrm{c}}{Q} = \mathrm{constant}$ and $\fn{P_\mathrm{c}}{Q} \propto Q^{-1}$.
In contrast to our previous model using a single mass constraint,
which gives a high typicality within $1 \sigma$,
this model may provide a high typicality, e.g.\ within $1 \sigma$,
or a low typicality, e.g.\ beyond $3 \sigma$,
depending on the mass history.
Note that these values are robust even if we apply the Tegmark \& Rees bound on $Q$ \cite{Tegmark:1997in}.

In \fig{fig:his_cont_lambda}, we compare the different multiverse measures for the typicality of $\Lambda_0$.
As seen in \fig{fig:his_measure},
in the case of the pocket based measure $\fn{P}{\Lambda|\smiley}$ mainly depends on the mass and time conditions.
On the other hand, in the cases of the scale factor cutoff and the causal patch measures, 
it mostly depends on the prior distribution from the measure itself.
As a result, both measures provide $\typicality{\Lambda_0}$ similar to that assuming only  $t_{\smiley} = t_0$.
Since their prior distributions are weighted toward $\Lambda \lesssim \Lambda_0$,
there even exists a mass history which gives $\typicality{\Lambda_0} = 1$.
However, along this mass history $\typicality{Q_0}$ is less than $10^{-5}$, i.e.\ beyond $3\sigma$, and this mass history is ruled out by $Q = Q_0$.

To determine whether models using mass history can actually help understanding $\Lambda_0$ and $Q_0$,
we make a quantitative example with a definite constraint.
To make an anthropic model we need to consider which historical factors may be anthropically important.
A galaxy may need to be sufficiently large at early times to produce or retain sufficient metals for life,
and it may also need to avoid dangerous interactions.
On the other hand, the galaxy may need to accrete sufficiently, 
for example, to stimulate star formation.

In order to make a quantitative model, 
we use observational studies in Ref.~\cite{Bullock:2005pi} that suggests that nearly 80\% of the current mass of the Milky Way came from an early major merger 10 billion years ago, 
and in Refs.~\cite{Hammer:2007ki,Burstein:2004pn} that suggest that there has not been any major interaction or a significant amount of minor mergers since then.
Interestingly, this is somewhat different to the history of Andromeda 
which may have experienced a more recent significant merger \cite{Burstein:2004pn}.
A comparative study of the merger histories and habitabilities of the Milky Way and Andromeda may be instructive.

Combining the above arguments, we consider the following toy models:
\paragraph{($M_\mathrm{i} \geq 0.8 M_\mathrm{MW}$ at $t_\mathrm{i} = 4\Gyr$) and ($M_\mathrm{f} = M_\mathrm{MW}$ at $t_\mathrm{f} = 6\Gyr$)}
to require that the galaxy was sufficiently large at a sufficiently early time.
\paragraph{($M_\mathrm{i} = 0.8 M_\mathrm{MW}$ at $t_\mathrm{i} = 4\Gyr$) and ($M_\mathrm{f} = M_\mathrm{MW}$ at $t_\mathrm{f} = 6\Gyr$)}
to require that the galaxy was sufficiently large at a sufficiently early time and had subsequent matter accretion.

\begin{table}[hbt]
\centering
\begin{tabular}{|c||c|c|c|c|}
\hline \multirow{2}{*}{$\typicality{\Lambda_0}$} & \multirow{2}{*}{PB} 
& PB & SFC & CP \\
& & $\fn{Q}{\Lambda,t_\mathrm{f}}$ & $\tau_{\smiley} = t_0$ & $\tau_{\smiley} = t_0$ \\[0.5ex]
\hline \hline $M_\mathrm{i} \geq 0.8 M_\mathrm{MW}$ & 0.011 & $2 \times 10^{-6}$ & 0.71 & 0.26 \\[0.5ex]
\hline $M_\mathrm{i} = 0.8 M_\mathrm{MW}$ & 0.045 & 0.023 & 0.55 & 0.25 \\[0.5ex]
\hline
\end{tabular}
\caption{ \label{tab:his_ex_lambda}
$\typicality{\Lambda_0}$ for the anthropic models with $t_\mathrm{i} = 4\Gyr$ and $M_\mathrm{f} = M_\mathrm{MW}$ at $t_\mathrm{f} = 6\Gyr$.
In the case $Q = \fn{Q}{\Lambda,t_\mathrm{f}}$ we take $\fn{P_\mathrm{c}}{Q} = \textrm{constant}$.
}
\end{table}

\begin{table}[hbt]
\centering
\begin{tabular}{|c||c|c|}
\hline $\typicality{Q_0}$ & $\fn{P_\mathrm{c}}{Q} = \mathrm{constant}$ & $\fn{P_\mathrm{c}}{Q} \propto Q^{-1}$ \\[0.5ex]
\hline \hline $M_\mathrm{i} \geq 0.8 M_\mathrm{MW}$ & 0.045 & 0.14 \\[0.5ex]
\hline $M_\mathrm{i} = 0.8 M_\mathrm{MW}$ & 0.41 & 0.67 \\[0.5ex]
\hline
\end{tabular}
\caption{ \label{tab:his_ex_Q}
$\typicality{Q_0}$ for the anthropic models with $t_\mathrm{i} = 4\Gyr$ and $M_\mathrm{f} = M_\mathrm{MW}$ at $t_\mathrm{f} = 6\Gyr$.
}
\end{table}

\twotab{tab:his_ex_lambda}{tab:his_ex_Q} show the typicalities of our toy models.
If we neglect the cases of the scale factor cutoff and the causal patch measures,
then the model $M_\mathrm{i} \geq 0.8 M_\mathrm{MW}$,
as might be suggested by the results of Refs.~\cite{Hammer:2007ki,Burstein:2004pn},
has some difficulty to explain both $\Lambda_0$ and $Q_0$.
On the other hand, in the case of the model $M_\mathrm{i} = 0.8 M_\mathrm{MW}$,
$\typicality{\Lambda_0}$ is greater, though maybe not sufficiently, than the case $M_\mathrm{i} \geq 0.8 M_\mathrm{MW}$,
and $\typicality{Q_0}$ is high, i.e.\ within $1 \sigma$.
Therefore, we set this model as our reference model.

\begin{table}[hbt]
\centering
\begin{tabular}{|c||c|c|c|c|}
\hline \multirow{2}{*}{$\typicality{\Lambda_0}$} & \multirow{2}{*}{PB} 
& PB & SFC & CP \\
& & $\fn{Q}{\Lambda,t_\mathrm{f}}$ & $\tau_{\smiley} = t_0$ & $\tau_{\smiley} = t_0$ \\[0.5ex]
\hline \hline $M_\mathrm{i} = 0.9 M_\mathrm{MW}$ at $t_\mathrm{i} = 3\Gyr$ & $4 \times 10^{-5}$ & $4 \times 10^{-6}$ & 0.36 & 0.48 \\[0.5ex]
\hline $M_\mathrm{i} = 0.8 M_\mathrm{MW}$ at $t_\mathrm{i} = 4\Gyr$ & 0.045 & 0.023 & 0.55 & 0.25 \\[0.5ex]
\hline $M_\mathrm{i} = 0.5 M_\mathrm{MW}$ at $t_\mathrm{i} = 5\Gyr$ & 0.10 & 0.077 & 0.59 & 0.22 \\[0.5ex]
\hline 
\end{tabular}
\caption{ \label{tab:his_ex2_lambda}
$\typicality{\Lambda_0}$ for the anthropic models with $M_\mathrm{f} = M_\mathrm{MW}$ at $t_\mathrm{f} = 6\Gyr$.
We slightly change the mass and time constraints from the model with $M_\mathrm{i} = 0.8 M_\mathrm{MW}$ at $t_\mathrm{i} = 4\Gyr$.
In the case $Q = \fn{Q}{\Lambda, t_\mathrm{f}}$ we take $\fn{P_\mathrm{c}}{Q} = \textrm{constant}$.
}
\end{table}

\begin{table}[hbt]
\centering
\begin{tabular}{|c||c|c|}
\hline $\typicality{Q_0}$ & $\fn{P_\mathrm{c}}{Q} = \mathrm{constant}$ & $\fn{P_\mathrm{c}}{Q} \propto Q^{-1}$ \\[0.5ex]
\hline \hline $M_\mathrm{i} = 0.9 M_\mathrm{MW}$ at $t_\mathrm{i} = 3\Gyr$ & $1 \times 10^{-4}$ & $4 \times 10^{-4}$ \\[0.5ex]
\hline $M_\mathrm{i} = 0.8 M_\mathrm{MW}$ at $t_\mathrm{i} = 4\Gyr$ & 0.41 & 0.67 \\[0.5ex]
\hline $M_\mathrm{i} = 0.5 M_\mathrm{MW}$ at $t_\mathrm{i} = 5\Gyr$ & 0.89 & 0.56\\[0.5ex]
\hline
\end{tabular}
\caption{ \label{tab:his_ex2_Q}
$\typicality{Q_0}$ for the anthropic models with $M_\mathrm{f} = M_\mathrm{MW}$ at $t_\mathrm{f} = 6\Gyr$.
We slightly change mass and time conditions from the model with $M_\mathrm{i} = 0.8 M_\mathrm{MW}$ at $t_\mathrm{i} = 4\Gyr$.
}
\end{table}

\twotab{tab:his_ex2_lambda}{tab:his_ex2_Q} show whether the results from our reference model, $M_\mathrm{i} = 0.8 M_\mathrm{MW}$ at $t_\mathrm{i} = 4\Gyr$, are robust even if we slightly change mass and time constraints.
In the case of $\typicality{\Lambda_0}$, 
the results from the scale factor cutoff measure and the causal patch measure are robust,
which only shows that $t_{\smiley} = t_0$ plays a more significant role than any other anthropic factor.
Note that the direction which increases the typicality for the scale factor cutoff measure is opposite to that for the causal patch measure.
On the other hand, in the cases of $\typicality{\Lambda_0}$ with the pocket based measure and $\typicality{Q_0}$,
the model with $M_\mathrm{i} = 0.9 M_\mathrm{MW}$ at $t_\mathrm{i} = 3\Gyr$ gives a low typicality.
Therefore, if the proper anthropic constraint consists of larger $M_\mathrm{i}$ and smaller $t_\mathrm{i}$ than our model,
the anthropic solution for both the cosmological constant and the primordial density perturbation amplitude would be in trouble,
and we may require additional anthropic constraints to solve this problem.

\section{Conclusion}
\label{sec:sum}

\begin{table}[hbt]
\centering
\begin{tabular}{|r @{~at~} l ||c|c|c|c|}
\hline \multicolumn{2}{|c||}{\multirow{2}{*}{$\typicality{\Lambda_0}$}} & \multirow{2}{*}{PB} 
& PB & SFC & CP \\
\multicolumn{2}{|c||}{} & & $\fn{Q}{\Lambda,t_\mathrm{f}}$ & $t_{\smiley} = t_0$ & $t_{\smiley} = t_0$ \\[0.5ex]
\hline \hline \multicolumn{2}{|c||}{-} & $2 \times 10^{-120}$ & $2 \times 10^{-15}$ & 0.55 & 0.14 \\[0.5ex]
\hline $M \geq M_\mathrm{MW}$ & $t \to \infty$ & 0.22 & $2 \times 10^{-15}$ & 0.36 & 0.11 \\[0.5ex]
\hline $M = M_\mathrm{MW}$ & $t = 6\Gyr$ & 0.049 & $2 \times 10^{-15}$ & 0.52 & 0.17 \\[0.5ex]
\hline $M = M_\mathrm{MW}$ & $t = 6\Gyr$ & \multirow{2}{*}{0.045} & \multirow{2}{*}{0.023} & \multirow{2}{*}{0.55} & \multirow{2}{*}{0.25} \\
$M = 0.8 M_\mathrm{MW}$ & $t = 4\Gyr$ & & & & \\[0.5ex]
\hline
\end{tabular}
\caption{\label{tab:sum_lambda}
Typicalities of $\Lambda_0$ for different anthropic models and multiverse measures:
the pocket based measure (PB), the scale factor cutoff measure (SFC) and the causal patch measure (CP).
We assume $0 \lesssim \Lambda \lesssim 1$ and $10^{-16} \lesssim Q \lesssim 1$.
$\fn{Q}{\Lambda,t_\mathrm{f}}$ makes the population of galaxies and clusters at $t = t_\mathrm{f}$ independent of $\Lambda$.
In the case $Q = \fn{Q}{\Lambda,t_\mathrm{f}}$, we take $\fn{P_\mathrm{c}}{Q} = \textrm{constant}$,
which gives the greatest difference to the case of $Q = Q_0$.
}
\end{table}

\begin{table}[hbt]
\centering
\begin{tabular}{|r @{~at~} l ||c|c|}
\hline \multicolumn{2}{|c||}{$\typicality{Q_0}$} & $\fn{P_\mathrm{c}}{Q} = \mathrm{constant}$ & $\fn{P_\mathrm{c}}{Q} \propto Q^{-1}$ \\[0.5ex]
\hline \hline \multicolumn{2}{|c||}{-} & $2 \times 10^{-5}$ & 0.63 \\[0.5ex]
\hline $M \geq M_\mathrm{MW}$ & $t \to \infty$ & $7 \times 10^{-7}$ & $8 \times 10^{-3}$ \\[0.5ex]
\hline $M = M_\mathrm{MW}$ & $t = 6\Gyr$ & 0.33 & 0.76 \\[0.5ex]
\hline $M = M_\mathrm{MW}$ & $t = 6\Gyr$ & \multirow{2}{*}{0.41} & \multirow{2}{*}{0.67} \\
$M = 0.8 M_\mathrm{MW}$ & $t = 4\Gyr$ & & \\[0.5ex]
\hline
\end{tabular}
\caption{\label{tab:sum_Q}
Typicalities of $Q_0$ for different anthropic models and prior distributions of $Q$, with $\Lambda = \Lambda_0$.
We assume $10^{-16} \lesssim Q \lesssim 1$.
}
\end{table}

We studied the comoving anthropic likelihood of an obsevable $O$, $\fn{L_\mathrm{c}}{O|\smiley}$, 
which counts the number of observers in a comoving volume, 
where $O$ corresponds to the cosmological constant $\Lambda$ and the primordial density perturbation amplitude $Q$.
To estimate $\fn{L_\mathrm{c}}{O|\smiley}$,
we started from Weinberg's anthropic calculation \cite{Martel:1997vi,Garriga:1999hu,Garriga:2000cv,Pogosian:2006fx}
which models the number of observers as proportional to the total mass in gravitationally collapsed objects 
with mass greater than a certain threshold, $M_*$, at late times, $t \to \infty$. 
While this model can postdict $\Lambda_0$ well with simple assumptions,
it assumes the supermassive objects are equally habitable to the Milky Way,
but they may not be very habitable while they give a bias to small $\Lambda$.
See the second row of \tab{tab:sum_lambda}.
Also, since supermassive objects prefer large $Q$,
Weinberg's model predicts large $Q$ unless one applies the Tegmark \& Rees' bound \cite{Tegmark:1997in} (see the second row of \tab{tab:sum_Q}).

In order to avoid the above problems of Weinberg's model, 
we considered a model that assumes that the number of observers is proportional to the number of gravitationally collapsed objects with certain mass and time scales $M = M_*$ and $t = t_*$.
Though it seems obvious to choose $M_*$ as the mass of the Milky Way and $t_*$ as $t_0$,
the Press-Schechter formalism \cite{Press:1973iz}, which we used to count the collapsed objects,
identifies our collapsed object at $t_0$ as the Local Group,
which makes it inconsistent to choose $M_*$ as the mass of the Milky Way.
Also, since we do not seem to have any plausible anthropic justification to use the Local Group as our object,
it is also inconsistent to choose $M_*$ as the mass of the Local Group.
However, the time before the formation of the Local Group may be anthropically more influential than $t_0$,
for example, by influencing the star formation rate or metal abundance, etc.
Also, if we set $t_*$ earlier than the formation of the Local Group,
we can identify the Milky Way as an isolated collapsed object
and the above technical problem with using the Press-Schechter formalism disappears.
Thus, we set $M_*$ as the mass of the Milky Way and $t_*$ as a time earlier than the formation of the Local Group, say, 6 billion years.
Since Weinberg's model was biased to small $\Lambda$,
the typicality of $\Lambda_0$ for our model in the pocket based measure is lower than Weinberg's by a factor of four.
See the third row of \tab{tab:sum_lambda}.
In the case of $Q$, our model can postdict $Q_0$ within $1\sigma$, while Weinberg's model predicts large $Q$
(see the third row of \tab{tab:sum_Q}).

Furthermore, it is not just the single mass constraint but the full mass history of a galaxy or a galaxy group which affects its habitability.
As a first step to consider the full mass history,
we introduced anthropic models assuming the number of observers is proportional to the number of gravitationally collapsed objects with $M=M_\mathrm{f}$ at $t = t_\mathrm{f}$,
which were formed from objects with $M = M_\mathrm{i}$ at an earlier time $t = t_\mathrm{i}$,
using the extended Press-Schechter formalism \cite{Bond:1990iw,Bower:1991kf,Lacey:1993iv}.
\figs{fig:his_cont_degeneracy}{fig:his_cont_lambda} show the typicalities of $\Lambda_0$ and $Q_0$ by choosing different $M_\mathrm{i}$ and $t_\mathrm{i}$ constraints and prior distributions.
Especially, as a toy model, we chose $M_\mathrm{i} = 0.8 M_\mathrm{MW}$ and $t_\mathrm{i} = 4 \Gyr$, 
since a galaxy may need to be a certain mass and mass fraction in earlier times
to produce sufficient metals and stimulate star formation, and also to avoid dangerous interactions.
Then the typicalities of both $\Lambda_0$ and $Q_0$ are similar to the model with the single mass constraint $M = M_\mathrm{MW}$ at $t = 6\Gyr$ (see the fourth row of \twotab{tab:sum_lambda}{tab:sum_Q}).
However, there is no degeneracy between $\Lambda$ and $Q$, 
which afflicts all kinds of single mass constraint models in the pocket based measure (see the second column of \tab{tab:sum_lambda}).

We also studied the effect of the multiverse measure on our typicality.
In addition to Weinberg/Vilenkin's flat prior/pocket based measure \cite{Garriga:1998px,Vanchurin:1999iv,Garriga:2005av,Easther:2005wi},
we considered two multiverse measures: the scale factor cutoff measure \cite{DeSimone:2008bq,Bousso:2008hz} and the causal patch measure \cite{Bousso:2006ev,Bousso:2006ge}.
In the case of the pocket based measure,
the typicality of $\Lambda_0$ is relatively small and sensitive to the choice of the anthropic model.
On the other hand, if we assume that the observing time $t_{\smiley} = t_0$,
both the scale factor cutoff measure and the causal patch measure 
always give a high typicality,
and it is not affected much by any other anthropic factors.

\begin{figure}[hbt]
\begin{center}
\includegraphics[height=0.3\textwidth]{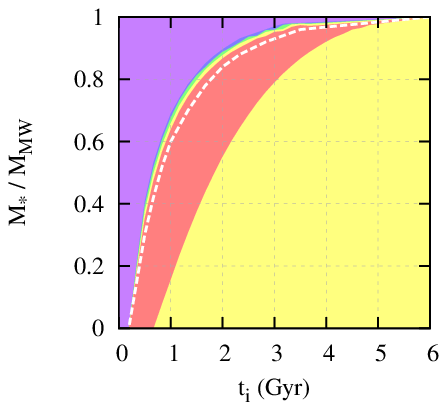} \hspace{20pt}
\includegraphics[height=0.3\textwidth]{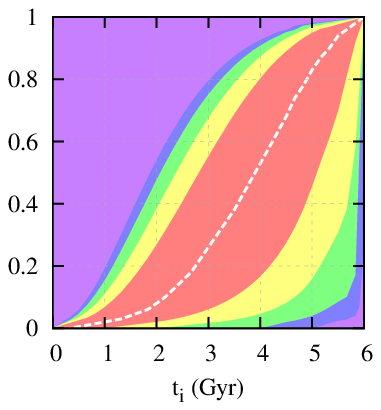}
\caption{\label{fig:mediocrity}
Examples which illustrate the difference between the mass history which \emph{makes our universe typical} and that which is \emph{typical in our universe}.
Left: the typicality of $\Lambda_0$ in the case of the causal patch measure.
Right: the typicality of mass history in our universe.
Typicality: {\color{violet} 0--0.01}, {\color{blue} 0.01--0.03}, {\color[rgb]{0,0.5,0} 0.03--0.1}, {\color[rgb]{0.4,0.4,0} 0.1--0.3}, {\color{red} 0.3--1}.
White dash: the maximal typicality.
}
\end{center}
\end{figure}

Note that one must be careful not to confuse the two separate questions:
whether a given mass history \emph{makes our universe typical}
and whether a given mass history is \emph{typical in our universe}.
There is a common misconception, called the ``Principle of Mediocrity,'' 
that our Galaxy is a typical galaxy in our universe, and our universe is a typical universe.
However, from the anthropic point of view, 
the typicality of $\Lambda_0$ from a given mass history and the typicality of that history in our universe are not expected to be similar.
For example, as shown in \fig{fig:mediocrity}, 
in the case of the causal patch measure,
the mass history which is most typical in our universe does not postdict $\Lambda_0$ within $1\sigma$,
and  the mass history which makes $\Lambda_0$ most typical is not typical in our universe within $2\sigma$.

\section{Discussion}\label{sec:dis}

The main problem of our work is how to choose anthropic factors or an anthropic model in terms of mass history.
Also, our calculation technique, the extended Press-Schechter formalism,
is crude and uses at most two historical points,
and it cannot follow the late history of a galaxy after it joins a galaxy group.

The actual history of the Milky Way can give hints for anthropic factors,
although one must be careful not to introduce bias by considering features due to our value of the cosmological constant,
as opposed to features affecting the formation of observers.
Interestingly, the history of the Milky Way seems to be somewhat different to the history of Andromeda.
From this we suggested that a comparative study of the habitabilities of the Milky Way and Andromeda may be instructive.

Cosmological numerical simulation may be able to consider the full history of a galaxy,
especially the late history.
These late times may provide the strongest anthropic constraint on the cosmological constant, 
since the effect of the cosmological constant on the large scale structure is greatest at late times.
This may give $\Lambda_0$ a high probability even in the case of the pocket based measure.
On the other hand, late times may not be so influential,
since the galaxy group may shield the effect of the cosmological constant.
This would support our previous argument that it may be better to set $t_\mathrm{f}$ as the time before the formation of the Local Group.

Up to now, we related the cosmological constant to the mass history,
which has only an indirect connection to the real anthropic factors.
What we actually need to do in the future is to relate the cosmological constant to physical properties those are more directly related to real anthropic factors, 
e.g.\ metallicity and star formation rate.
Numerical simulation may help to make this possible.

\section*{Acknowledgments}
The authors thank  Changbom Park, Jai-chan Hwang, Juhan Kim, Donghui Jeong, Jinn-Ouk Gong,
 Kenji Kadota,   Dong-han Yeom, Seoktae Koh, Cai Kai, Michael Gowanlock, Alex Nielson, Bum-Hoon Lee
and Emanuil Vilkovisky.
The authors also thank the hospitality of Center for Quantum SpaceTime, Fesenkov Astrophysical Institute, Korea Institute for Advanced Study and Yukawa Institute of Theoretical Physics.
SEH and EDS are supported by the National Research Foundation grant (2009-0077503) funded by the Korean government.
SEH is also supported by the National Research Foundation grant (2009-006814, 2007-0093860) funded by the Korean government.
HZ is supported by T\"UB\.ITAK research fellowship programme for foreign citizens.

\newpage
\appendix

\section{Prior distribution of $\Lambda$ for different measures}\label{app:prior}

\subsection{Scale factor cutoff measure}\label{app:prior_sfc}

The probability of the universe with the age $t_{\smiley}$ and the cosmological constant $\Lambda$
using the scale factor cutoff measure \cite{DeSimone:2008bq,Bousso:2008hz}
is proportional to the limit of the thermalized volume using a certain scale factor cutoff $a_\mathrm{c}$:
\begin{equation}\label{eq:sfc}
\fn{W}{\Lambda, t_{\smiley}} \propto
\lim_{a_\mathrm{c} \to \infty} \fn{a^3}{\Lambda,\fn{t}{a_\mathrm{c},\Lambda} - t_{\smiley}} ,
\end{equation}
where 
$\fn{a}{\Lambda,t}$ is the scale factor of the universe with the cosmological constant $\Lambda$ at $t$,
$\fn{t}{a,\Lambda}$ is the time when the scale factor of the universe with the cosmological constant $\Lambda$ becomes $a$.
Here we consider the volume which thermalizes at $t_\mathrm{therm} = \fn{t}{a_\mathrm{c},\Lambda} - t_{\smiley}$
so that the age of the thermalized region should be $\fn{t}{a_\mathrm{c},\Lambda} - t_\mathrm{therm} = t_{\smiley}$.

If $t \gg 1$, then the universe with $\Lambda > 0$ can be approximated as a de Sitter space, so
\begin{equation}
\fn{a}{\Lambda,t} \sim \exp \left\{ \sqrt{\frac{8 \pi G \Lambda}{3}} t \right\} .
\end{equation}
Assuming $\fn{t}{a_\mathrm{c},\Lambda} \gg 1$, which is proper since we will later use $a_\mathrm{c} \to \infty$,
\eq{eq:sfc} becomes
\begin{equation}
\fn{W}{\Lambda,t_{\smiley}} \propto \exp \left\{ -2.55 \sqrt{\frac{\Lambda}{\Lambda_0}} \frac{t_{\smiley}}{t_0} \right\} .
\end{equation}

\subsection{Causal patch measure}\label{app:prior_cp}
The probability of the universe with the age $t_{\smiley}$ and the cosmological constant $\Lambda$
using the causal patch measure \cite{Bousso:2006ev,Bousso:2006ge}
is proportional to the volume of the causally connected region:
\begin{equation}
\fn{W}{\Lambda,t_{\smiley}} \propto \min \left[ \fn{\tau^3}{\Lambda,t_{\smiley}}, 
(\fn{\tau}{\Lambda,\infty} - \fn{\tau}{\Lambda,t_{\smiley}})^3 \right] ,
\end{equation}
where
\begin{equation}
\fn{\tau}{\Lambda,t} \equiv \int_0 ^t \frac{\d{t'}}{\fn{a}{\Lambda,t'}}
\propto \left( \frac{\Lambda}{\Lambda_0} \right)^{\frac{1}{3}} \int_0 ^t \d{t'} \sinh^{-\frac{2}{3}} \left\{1.29 \sqrt{\frac{\Lambda}{\Lambda_0}} \frac{t'}{t_0} \right\}
\end{equation}
is the conformal time of the universe with $\Lambda$ at $t$.

\section{Smoothed density field}\label{app:sigma}
\subsection{$Q = Q_0$}\label{app:sigma_q0}
The current smoothed density field of our universe $\fn{\sigma}{M}$ is calculated \cite{Spergel:2006hy} from the matter power spectrum $\fn{P}{k}$,
\begin{equation}
\fn{\sigma}{M} = \left\{ \frac{1}{2 \pi ^2} \int_0 ^{\infty} {\d{k} \fn{P}{k} \fn{W^2}{k \left[\frac{3M}{4 \pi \rho_\mathrm{m}}\right]^{\frac{1}{3}} } k^2} \right\} ^{\frac{1}{2}}.
\end{equation}
with the top-hat filter 
\begin{equation}
\fn{W}{kR} \equiv \frac{3 (\sin kR - kR \cos kR)}{(kR)^3}.
\end{equation}
In universes where $\Lambda \lesssim 10^3 \Lambda_0$,
the effect from the cosmological constant is negligible at the recombination era ($z \sim 1000$).
If we assume that the density perturbation of all universes at that time are identical,
then we can calculate the smoothed density field $\fn{\sigma}{M,t;\Lambda}$ from the growth function $\fn{D}{t;\Lambda}$,
\begin{equation}
\frac{\fn{\sigma}{M,t;\Lambda}}{\fn{D}{t;\Lambda}} = \frac{\fn{\sigma}{M}}{\fn{D}{t_0;\Lambda_0}}\,.
\end{equation}
The analytic form of the linear growth function is \cite{Martel:1991}
\begin{equation}
\fn{D}{t;\Lambda} \simeq \frac{3}{2} \left( \frac{\fn{\rho_{\Lambda}}{t_\mathrm{rec};\Lambda}}{\fn{\rho_\mathrm{m}}{t_\mathrm{rec};\Lambda}} \right)^{-\frac{1}{3}} \fn{G}{ \frac{\fn{\rho_{\Lambda}}{t;\Lambda}}{\fn{\rho_\mathrm{m}}{t;\Lambda}} }, \label{eq:app_d_second}
\end{equation}
where $t_\mathrm{rec}$ is the physical time at the recombination
and
\begin{equation}
\fn{G}{x} \equiv \frac{5}{6} \sqrt{ \frac{1+x}{x} } \int_{0} ^{x} 
\frac{\d{w}} {w^{\frac{1}{6}} (1+w)^{\,\frac{3}{2}}}\,.
\end{equation}
Therefore,
\begin{align}
\fn{\sigma}{M,t;\Lambda} &= \fn{\sigma}{M} \frac{\fn{D}{t; \Lambda} }{\fn{D}{t_0 ; \Lambda_0 } } \\
&= 0.93\, \fn{\sigma}{M} \left(\frac{\Lambda}{\Lambda_0}\right)^{-\frac{1}{3}} 
\fn{G}{ \sinh^2 \left\{ 1.29 \sqrt{\frac{\Lambda}{\Lambda_0}} \frac{t}{t_0} \right\} \!\! } .
\end{align}

\subsection{$Q = \fn{Q}{\Lambda,t_*}$} \label{app:sigma_qfunc}
We define $Q = \fn{Q}{\Lambda,t_*}$ to satisfy
\begin{equation}
\fn{\sigma}{M,t_*;\Lambda,\fn{Q}{\Lambda,t_*}} = \fn{\sigma}{M,t_*;\Lambda_0,Q_0}. \label{eq:q_def}
\end{equation} 
Since the smoothed density field is proportional to the primordial density perturbation amplitude,
\begin{align}
\fn{\sigma}{M,t;\Lambda,Q} &\equiv \frac{Q}{Q_0} \fn{\sigma}{M,t;\Lambda,Q_0} \\
&= \fn{\sigma}{M} \frac{Q}{Q_0} \frac{\fn{D}{t;\Lambda}}{\fn{D}{t_0;\Lambda_0}}. \label{eq:q_def2}
\end{align}
Then, using \twoeq{eq:q_def}{eq:q_def2},
\begin{align}
\fn{Q}{\Lambda,t_*}
&= Q_0 \frac{ \fn{D}{t_*; \Lambda_0} }{\fn{D}{t_*; \Lambda}} \\
&= Q_0 \left(\frac{\Lambda}{\Lambda_0}\right)^{\frac{1}{3}}
\frac{\displaystyle \fn{G}{ \sinh^2 \left\{ 1.29 \,\,\frac{t_*}{t_0} \right\} \!\! }}
{\displaystyle \fn{G}{ \sinh^2 \left\{ 1.29 \sqrt{\frac{\Lambda}{\Lambda_0}} \frac{t_*}{t_0} \right\} \!\!}} ,
\end{align}
and the corresponding smoothed density field is
\begin{align}
\fn{\sigma}{M,t;\Lambda,\fn{Q}{\Lambda,t_*}} 
&= \fn{\sigma}{M} \frac{\fn{Q}{\Lambda, t_*} }{Q_0} \frac{\fn{D}{t; \Lambda}}{\fn{D}{t_0; \Lambda_0}} \\
&= 0.93 \fn{\sigma}{M} \fn{G}{ \sinh^2\left\{ 1.29 \,\,\frac{t_*}{t_0} \right\} \!\! }
\frac{\displaystyle \fn{G}{\sinh^2\left\{1.29 \sqrt{\frac{\Lambda}{\Lambda_0}} \frac{t}{t_0} \right\}\!\!} }
{\displaystyle \fn{G}{\sinh^2 \left\{1.29 \sqrt{\frac{\Lambda}{\Lambda_0}} \frac{t_*}{t_0} \right\} \!\!}} . \label{eq:q_sigma}
\end{align}

\section{Calculation techniques using single mass constraint}\label{app:final}
\subsection{$\fn{L_\mathrm{c}}{\Lambda|\smiley}$} \label{app:final_lambda}
From the Press-Schechter formalism \cite{Press:1973iz}, 
the mass fraction of the gravitationally collapsed objects whose mass is greater than $M$ at time $t$ is
\begin{equation} \label{eq:classic_frac}
\int_{M}^{\infty} \d{M} \fn{F_1}{M,t} = \frac{2}{\sqrt{2\pi}\, \fn{\sigma}{M,t}}
\int_{\delta_\mathrm{c}}^\infty \d\delta \exp\left\{ -\frac{\delta^2}{2\fn{\sigma^2}{M,t}} \right\} \,, 
\end{equation}
where $\delta_\mathrm{c} \sim \mathinner{1.68}$ is the critical collapse overdensity.
Then the anthropic likelihood for the model $M \geq M_*$ at $t \to \infty$ is proportional to \eq{eq:classic_frac}, 
\begin{equation}
\fn{L_\mathrm{c}}{\Lambda| \smiley, M \geq M_*, t \to \infty}
\varpropto \erfc \left\{ \frac{1}{\sqrt{2}} \frac{\delta_\mathrm{c}}{\fn{\sigma}{M_*, \infty; \Lambda}} \right\}. \label{eq:classic_dist}
\end{equation}

By differentiating \eq{eq:classic_frac},
one can obtain the mass fraction of the gravitationally collapsed objects with mass $[M,M+\d{M}]$ at $t$,
\begin{equation}
\fn{F_1}{M, t} \, \d{M} 
= \frac{1}{\sqrt{2\pi}} \frac{\delta_\mathrm{c}}{\fn{\sigma}{M,t}} 
\exp \left\{ -\frac{\delta^2_\mathrm{c}} {2 \fn{\sigma^2}{M,t}} \right\}
\frac{\d{\fn{\sigma^2}{M,t}}}{\fn{\sigma^2}{M,t}}\,. \label{eq:mass_frac}
\end{equation}
Then the anthropic likelihood for the model $M = M_*$ at $t = t_*$ is proportional to \eq{eq:mass_frac},
\begin{equation}
\fn{L_\mathrm{c}}{\Lambda|\smiley,M=M_*, t = t_*}
\varpropto \frac{\delta_\mathrm{c}}{\fn{\sigma}{M_*,t_*; \Lambda}} \exp \left\{ -\frac{\delta_\mathrm{c} ^2}{2 \fn{\sigma^2}{M_*,t_*; \Lambda}} \right\}. \label{eq:mass_dist}
\end{equation}

\subsection{$\fn{L_\mathrm{c}}{Q|\smiley}$}\label{app:final_Q}
By fixing $\Lambda = \Lambda_0$ and combining \twoeq{eq:classic_frac}{eq:mass_frac} with \eq{eq:q_def2},
the anthropic likelihood of $Q$ for the model $M \geq M_*$ at $t \to \infty$ is
\begin{equation}
\fn{L_\mathrm{c}}{Q| \smiley, M \geq M_*, t \to \infty}
\varpropto \left( \frac{Q}{Q_0} \right)^{-1} \erfc \left\{ \frac{1}{\sqrt{2}} \frac{\delta_\mathrm{c}}{\fn{\sigma}{M_*, \infty}} \left( \frac{Q}{Q_0} \right)^{-1} \right\},
\end{equation}
and the anthropic likelihood of $Q$ for the model $M = M_*$ at $t = t_*$ is
\begin{equation}
\fn{L_\mathrm{c}}{Q|\smiley,M=M_*, t = t_*}
\varpropto \left( \frac{Q}{Q_0} \right)^{-2} \exp \left\{ -\frac{\delta_\mathrm{c} ^2}{2 \fn{\sigma^2}{M_*,t_*}} \left( \frac{Q}{Q_0} \right)^{-2} \right\}.
\end{equation}

\section{Calculation techniques using mass history}
\label{app:history}
\subsection{$\fn{L_\mathrm{c}}{\Lambda|\smiley}$ for the case of $Q = Q_0$}\label{app:history_lambda_q0}
The extended Press-Schechter formalism \cite{Bond:1990iw,Bower:1991kf,Lacey:1993iv} calculates the mass fraction of collapsed objects 
whose mass was $M_\mathrm{i}$, from objects whose mass is $M_\mathrm{f}$:
\begin{equation}
\begin{split}
\fn{F_2}{M_\mathrm{i},t_\mathrm{i} | M_\mathrm{f},t_\mathrm{f}}\, \d{M_\mathrm{i}} &= \frac{1}{\sqrt{2\pi}} \frac{\fn{\delta}{t_\mathrm{i}}-\fn{\delta}{t_\mathrm{f}}}{\sqrt{\fn{\sigma^2}{M_\mathrm{i}} - \fn{\sigma^2}{M_\mathrm{f}}}} \\
\times& \exp \left\{ -\frac{\left[\fn{\delta}{t_\mathrm{i}}-\fn{\delta}{t_\mathrm{f}}\right]^2}{2\left[\fn{\sigma^2}{M_\mathrm{i}} - \fn{\sigma^2}{M_\mathrm{f}}\right]} \right\} 
\frac{\d{\left[ \fn{\sigma^2}{M_\mathrm{i}} - \fn{\sigma^2}{M_\mathrm{f}} \right]}}{\fn{\sigma^2}{M_\mathrm{i}} - \fn{\sigma^2}{M_\mathrm{f}}}\,,
\end{split}
\label{eq:app_f2}
\end{equation}
where $\fn{\delta}{t} \equiv \delta_\mathrm{c}\, {\fn{D}{t_0}}/{\fn{D}{t}}$ and $\fn{\sigma}{M} \equiv \fn{\sigma}{M,t_0}$.
If the earlier mass constraint is fixed, i.e.\ $M_\mathrm{i} = M_*$, 
the anthropic likelihood of $\Lambda$ is proportional to $\fn{F_1}{M_\mathrm{f},t_\mathrm{f}} \, \fn{F_2}{M_*,t_\mathrm{i}|M_\mathrm{f},t_\mathrm{f}}$,
\begin{equation}
\begin{split}
\fn{L_\mathrm{c}}{\Lambda|\smiley,M_\mathrm{i} = M_*, Q_0}
&\varpropto \frac{\fn{\delta}{t_\mathrm{f};\Lambda}}{\fn{\sigma}{M_\mathrm{f}}} 
\exp \left( -\frac{\fn{\delta^2}{t_\mathrm{f};\Lambda}} {2 \fn{\sigma^2}{M_\mathrm{f}}} \right) \\
\times & \frac{\fn{\delta}{t_\mathrm{i}; \Lambda} - \fn{\delta}{t_\mathrm{f};\Lambda}}{\sqrt{\fn{\sigma^2}{M_*} - \fn{\sigma^2}{M_\mathrm{f}}}}
\,\exp \left\{ -\frac{\left\{\fn{\delta}{t_\mathrm{i}; \Lambda} - \fn{\delta}{t_0;\Lambda}\right\}^2} {2 \left[\fn{\sigma^2}{M_*} - \fn{\sigma^2}{M_\mathrm{f}}\right]} \right\}.
\end{split}
\end{equation}

By integrating $F_2$ from $M_*$ to $M_\mathrm{f}$, we can also calculate the anthropic likelihood of $\Lambda$ for the case $M_\mathrm{i} \geq M_*$:
\begin{equation}
\begin{split}
\,&\fn{L_\mathrm{c}}{\Lambda|\smiley,M_\mathrm{i} \geq M_*, Q_0} \\
\,&~~\varpropto \frac{\fn{\delta}{t_\mathrm{f};\Lambda}}{\fn{\sigma}{M_\mathrm{f}}} 
\exp \left\{ -\frac{\fn{\delta^2}{t_\mathrm{f};\Lambda}}{2\fn{\sigma^2}{M_\mathrm{f}}} \right\}
\erfc \left\{ \frac{1}{\sqrt{2}}\frac{\fn{\delta}{t_\mathrm{i};\Lambda} - \fn{\delta}{t_\mathrm{f};\Lambda}}{\sqrt{\fn{\sigma^2}{M_*} - \fn{\sigma^2}{M_\mathrm{f}}}} \right\}.
\end{split}
\end{equation}
In the similar way, we can also calculate the anthropic likelihood of $\Lambda$ for $M_\mathrm{i} \leq M_*$:
\begin{equation}
\begin{split}
\,&\fn{L_\mathrm{c}}{\Lambda|\smiley,M_\mathrm{i} \leq M_*, Q_0} \\
\,&~~\varpropto \frac{\fn{\delta}{t_\mathrm{f};\Lambda}}{\fn{\sigma}{M_\mathrm{f}}} 
\exp \left\{ -\frac{\fn{\delta^2}{t_\mathrm{f};\Lambda}}{2\fn{\sigma^2}{M_\mathrm{f}}} \right\} 
\erf \left\{ \frac{1}{\sqrt{2}}\frac{\fn{\delta}{t_\mathrm{i};\Lambda} - \fn{\delta}{t_\mathrm{f};\Lambda}}{\sqrt{\fn{\sigma^2}{M_*} - \fn{\sigma^2}{M_\mathrm{f}}}} \right\}.
\end{split}
\end{equation}

\subsection{$\fn{L_\mathrm{c}}{\Lambda|\smiley}$ for the case of $Q = \fn{Q}{\Lambda,t_\mathrm{f}}$}\label{app:history_lambda_qfunc}
Compared to the results from Appendix~\ref{app:history_lambda_q0},
there are two differences:
\begin{enumerate}
\item From \eq{eq:q_sigma},
all $\fn{\delta}{t;\Lambda}$ terms are replaced to 
\begin{equation}
\fn{\delta}{t;\Lambda} \times \left(\frac{\Lambda}{\Lambda_0}\right)^{-\frac{1}{3}} 
\frac{\displaystyle \fn{G}{ \sinh^2 \left\{1.29 \sqrt{\frac{\Lambda}{\Lambda_0}} \frac{t_\mathrm{f}}{t_0}\right\}\!\!}}
{\displaystyle \fn{G}{ \sinh^2 \left\{1.29\,\, \frac{t_\mathrm{f}}{t_0} \right\}\!\!}}.
\end{equation}
\item All terms related to $\fn{F_1}{M,t}$ are replaced to $\fn{P}{\fn{Q}{\Lambda,t_\mathrm{f}}}$.
\end{enumerate}

\subsection{$\fn{L_\mathrm{c}}{Q|\smiley}$}\label{app:history_Q}
Compared to the results from Appendix~\ref{app:history_lambda_q0},
there are two differences:
\begin{enumerate}
\item Since we fix $\Lambda = \Lambda_0$,
all $\fn{\delta}{t;\Lambda}$ terms are replaced to $\fn{\delta}{t}$.
\item From \eq{eq:q_def2},
all $\fn{\sigma}{M}$ terms are replaced to $\displaystyle \fn{\sigma}{M} \frac{Q}{Q_0}$.
\end{enumerate}

\end{document}